\documentclass{jfm}
\usepackage{graphicx}
\usepackage{epstopdf, epsfig}
\usepackage{newtxtext}
\usepackage{newtxmath}
\usepackage{natbib}
\usepackage{hyperref}
\hypersetup{
	colorlinks = true,
	urlcolor   = blue,
	citecolor  = black,
}
\usepackage[x11names, svgnames, dvipsnames]{xcolor}
\usepackage{soul} 
\usepackage{float}
\usepackage{lineno}
\usepackage{physics}
\usepackage[normalem]{ulem}

\newcommand{\St}{\mbox{\it St}}

\makeatletter
\def\sgn{\mathop{\operator@font sgn}}
\makeatother

\hypersetup{
	colorlinks = true,
	urlcolor   = blue,
	citecolor  = black,
}

\title{Large-scale streaks in a turbulent bluff body wake}
\shorttitle{Streaks in a turbulent wake}	

\author{Akhil Nekkanti\aff{1} \corresp{Present address: Division of Engineering and Applied Science, California Institute of Technology,
Pasadena, CA 91125, USA }, Sheel Nidhan\aff{1}, Oliver T. Schmidt\aff{1} \and Sutanu Sarkar\aff{1}  \corresp{\email{sarkar@ucsd.edu}}}

\affiliation{\aff{1}Department of Mechanical and Aerospace Engineering, University of California San Diego, CA 92093, USA}

\begin{document}

\maketitle

\begin{abstract}
A turbulent circular disk wake database (Chongsiripinyo \& Sarkar, \textit{J. Fluid Mech.}, vol. 885, 2020) at Reynolds number $\Rey = U_\infty D/\nu = 5 \times 10^{4}$ is interrogated to  identify  the presence of large-scale streaks - coherent elongated regions of streamwise velocity. The unprecedented streamwise length - until $x/D \approx 120$ - of the simulation enables investigation of the near and  far wake. The near wake  is dominated by the vortex shedding (VS) mode residing at azimuthal wavenumber $m=1$ and Strouhal number $\St = 0.135$. After filtering out the VS structure,  conclusive evidence of large-scale streaks with frequency $\St \rightarrow 0$,  equivalently streamwise wavenumber $k_x \rightarrow 0$ in the wake, becomes apparent in visualizations and spectra. These streaky structures are found  throughout the simulation domain beyond $x/D \approx 10$. Conditionally averaged streamwise vorticity fields reveal that the lift-up mechanism is active in the near as well as the far wake, and  that ejections contribute more than sweeps to events of 
 intense $-u'_xu'_r$.
Spectral proper orthogonal decomposition (SPOD) is employed to extract the energy and the spatiotemporal features of the large-scale streaks. The streak energy is concentrated in the $m=2$ azimuthal mode over the entire domain. Finally, bispectral mode decomposition (BMD) is conducted to reveal
strong interaction between $m=1$ and $\St = \pm 0.135$ modes to give the $m=2, \St \rightarrow 0$ streak mode. Our results indicate that the self-interaction of the VS mode generates the $m=2, \St \rightarrow 0$ streamwise vortices, which leads to streak formation through the lift-up process. To the authors' knowledge, this is the first study that reports and characterizes large-scale low-frequency streaks and the associated lift-up mechanism in a turbulent wake.  
\end{abstract}

\begin{keywords} 
\end{keywords}

\section{Introduction}\label{sec:introdcution}

Coherent structures, which are organized patterns of motion in a seemingly random turbulent flow field, play an essential role in  turbulent  shear flows.  Among these structures, streaks are among the most widely discussed, particularly in wall-bounded flows, where they were identified experimentally \citep{kline1967structure} as elongated regions of streamwise velocity in the near-wall region. As these streaks break up, they transfer energy from the inner to the outer layers, thereby maintaining turbulence in the outer layers of the boundary layer \citep{kim1971production}. This process, also known as \emph{bursting}, can account for up to 75\% of Reynolds stresses \citep{lu1973measurements}, hence assisting in the production of turbulent kinetic energy (TKE). \citet{smith1983characteristics} found that  low-speed streaks are robust features of boundary layers, occurring across a wide range of Reynolds numbers ($740 < \Rey_\theta < 5830$). Their spanwise spacing of 100 wall units was found to be invariant with $\Rey_\theta$. \citet{hutchins2007evidence} investigated the logarithmic region of the boundary layer, finding that the streaks in this region are distinct and much larger than the near-wall streaks, extending up to 20 times  the boundary layer thickness. Building upon this work,  \cite{monty2007large} reported the existence of streaks in the logarithmic region of turbulent pipe and channel flows as well. They also found that the width of these structures in  channel and pipe flows is larger than those of the boundary layer.  

The presence of a wall is not a prerequisite for the formation of streaks \citep{jimenez1999autonomous,mizuno2013wall}. A few studies in the past
 \citep[see figure 8b][]{brown1974density,bernal1986streamwise, liepmann_role_1992} have reported the presence of streak-like structures in
 the mixing layer, with the latter two  showing increasing  amplification of the streaks as the flow progresses downstream. More attention has been paid to the role of streaks in the mixing layer and specially the  jet, recently. \citet{jimenez-gonzalez_transient_2017} performed transient growth analysis in round jets finding that, for optimal initial disturbances, the streamwise vortices evolve to produce streamwise streaks. \citet{marant_influence_2018} also reported similar findings in a hyperbolic-tangent mixing layer. \citet{nogueira_large-scale_2019} applied spectral proper orthogonal decomposition on a PIV dataset of a circular turbulent jet at a high Reynolds number and demonstrated the presence of large-scale streaky structures in the near field (until $x/D = 8$). They further demonstrated that these structures exhibit large time scales and are associated with a low frequency of $\St \rightarrow 0$. The numerical counterpart of the previous study was performed by \citet{pickering_lift-up_2020}. They found that the streaky structures near the nozzle exit are dominated by higher azimuthal wavenumbers and the dominance shifts to lower azimuthal wavenumbers downstream, with $m=2$ dominating by $x/D = 30$. A similar conclusion about the  dominance of $m=2$ was reached by \citet{samie_three-dimensional_2022} who utilized quadrant analysis on a low Reynolds number jet. 

 In wall-bounded flows, streaks are generated by the lift-up mechanism \citep{ellingsen_stability_1975, landahl1975wave}.   Streamwise vortices  induce wall-normal velocities, bringing fluid from high-speed to low-speed regions and vice-versa to form streaks, hence the term `lift-up'. The subsequent instability and breakdown of these streaks are important to the self-sustaining cycle of wall turbulence \citep{hamilton1995regeneration, waleffe1997self}. \cite{brandt2014lift} presents a detailed review of the theory behind the lift-up mechanism and its role in transitional and turbulent flows. Although originally introduced as an instability mechanism that destabilizes a streamwise-independent base flow, the lift-up mechanism has been found to be dynamically crucial to fully turbulent wall-bounded flows as well \citep{jimenez2018coherent, bae2021nonlinear, farrell2012dynamics}.

The lift-up mechanism is active and plays a critical role in jets too. In their resolvent analysis of data from a turbulent jet experiment,  \cite{nogueira_large-scale_2019} found that the optimal forcing modes at $\St \rightarrow 0$ take the form of streamwise vortices that eject high-speed fluid and sweep low-speed fluid, depending on the orientation of these vortices. \cite{pickering_lift-up_2020}  analyzed a  turbulent jet LES database, finding that the response modes of these lift-up dominated optimal forcing modes indeed take the form of streamwise streaks. While these studies focused on circular jets, \cite{lasagna_near-field_2021} found that the lift-up mechanism is active in the near field of fractal jets as well.

As discussed in the foregoing, there has been a growing interest in the investigation of streaky structures and lift-up mechanisms in free shear flows, particularly in turbulent jets. These experimental and numerical studies have confirmed that the presence of a wall is not necessary for the formation of streaks, which has motivated us to explore another important class of free shear flows, i.e., turbulent wakes. Previous wake studies
 have primarily focused on the vortex shedding mechanism. Near the body and in the intermediate wake, the vortex shedding (VS) mode emerges as the most dominant coherent structure \citep{taneda1978visual,cannon1993observations,berger1990coherent, yun2006vortical}. However, it is worth noting that \cite{johansson_proper_2002} reported the presence of a distinct very low-frequency mode, $\St \rightarrow 0$, at azimuthal wavenumber $m=2$ in their proper orthogonal decomposition (POD) analyses of the turbulent wake of a circular disk at $\Rey \approx 2.5 \times 10^4$. This mode is distinct from the VS mode of a circular disk wake, which resides at $m=1$ with $\St=0.135$ \citep{berger1990coherent}. In a subsequent study, \cite{johansson_far_2006} extended the downstream distance to $x/D = 150$ and found that the $m=2$ mode with $\St \rightarrow 0$ dominated the far wake of the disk in terms of energy content relative to the vortex shedding mode. This finding was  corroborated in the spectral POD analysis \citep{nidhan2020spectral} of a disk wake at a higher Reynolds number ($\Rey = 5 \times 10^4$). The authors further found  low-rank behavior of the SPOD modes and that almost the entire Reynolds shear stress could be reconstructed with the leading SPOD modes of $m \leq 4$. Streamwise-elongated streaks, a focus of the current paper, were not considered by \cite{nidhan2020spectral}.  

None of the wake studies that report the presence and the eventual dominance at large $x$ of the very low-frequency mode ($\St \rightarrow 0$) at $m=2$ explain the physical origins of this structure. To address this gap, we revisit the LES database of \cite{chongsiripinyo_decay_2020}, 
who simulated the wake of a circular disk up to an unprecedented downstream distance of $x/D =125$. The large streamwise domain enables us to investigate  the entire wake. Unlike the previous SPOD  analysis~\citep{nidhan2020spectral} of this wake database, we focus on the streaky structures. We attempt to answer the following questions: Can  streaky structures  be identified in the near and far field of the turbulent wake? Is the lift-up mechanism active in the turbulent wake? How do the energetics and spatial structure of streaks evolve with downstream distance? What, if any, is the link  between the streak and the well-documented and extensively studied vortex shedding mode? 

In this work, besides visualizations and classical statistical analyses, we utilize two modal techniques, spectral proper orthogonal decomposition (SPOD) and bispectral mode decomposition (BMD), to shed light on the aforementioned questions. SPOD, whose mathematical framework in the context of turbulent flow was laid out by  \citet{lumley_1967,lumley_1970}, extracts a set of orthogonal modes sorted according to their energy at each frequency. It distinguishes the different time scales of the flow and identifies the most energetic coherent structures at each time scale. SPOD modes are coherent in both space and time and represent the flow structures in a statistical sense \citep{towne_spectral_2018}. Early applications of SPOD were by \cite{glauser1987coherent,glauser1992application,delville1994characterization} and this method has regained popularity since the work by \citet{towne_spectral_2018,schmidt_spectral_2018}. SPOD is particularly suitable for detecting and educing modes corresponding to streaky structures in statistically stationary flows \citep{nogueira_large-scale_2019, pickering_lift-up_2020, abreu_spectral_2020}, and is hence employed in this work. Bispectral mode decomposition, proposed by \citet{schmidt2020bispectral}, extracts the flow structures that are generated through triadic interactions. It identifies the most dominant triads in the flow by maximizing the spatially-integrated bispectrum. Furthermore, it picks out the spatial regions of nonlinear interactions between the coherent structures. BMD has been used to characterize the triadic interactions in various flow configurations, such as laminar-turbulent transition on a flat plate \citep{goparaju2022role}, forced jets \citep{maia2021nonlinear,nekkanti2022triadic,nekkanti2023bispectral}, swirling flows \citep{moczarski2022interaction} and wake of an airfoil \citep{patel2023modal}. In this work, we will employ BMD to investigate the presence and strength of nonlinear interactions between the VS mode and the streak-containing modes.

The manuscript is organized as follows. In \S\ref{sec:methodology_database},  the dataset and numerical methodology of SPOD and BMD are discussed. \S \ref{sec:visualizations} presents the extraction and visualization of streaks and lift-up mechanism in the near and far wake. Results from SPOD analysis at different downstream locations are presented in \S\ref{sec:spod_analysis} with a particular emphasis again on streaks and lift-up mechanism. \S\ref{sec:nonlinear_interaction} presents the results from the analysis of nonlinear interactions in the wake at select locations. The manuscript ends  with discussion and conclusions in \S \ref{sec:conclusions}.

\section{
Numerical wake database and methods for its analysis
}\label{sec:methodology_database}

\subsection{Dataset description}
The dataset employed for the present study of wake dynamics is from the  high-resolution large eddy simulation (LES)  of flow past a circular disk at  Reynolds number, $\Rey = U_\infty D/\nu = 5 \times 10^4$, reported in \cite{chongsiripinyo_decay_2020}. Here $U_\infty$ is the freestream velocity, $D$ is the disk diameter, and $\nu$ is the kinematic viscosity. The case of a homogeneous fluid  from \cite{chongsiripinyo_decay_2020}, who also simulate stratified wakes, is selected here.  The filtered Navier-Stokes equations, subject to the  condition of solenoidal velocity, were numerically solved on a structured cylindrical grid that spans a radial distance of $r/D = 15$ and a streamwise distance of $x/D = 125$. An immersed boundary method \citep{balaras_modeling_2004} is used to represent the circular disk in the simulation domain and the dynamic Smagorinsky model \citep{germano_dynamic_1991} is used for the LES model. The number of grid points in the radial ($r$), azimuthal ($\theta$) and streamwise directions ($x$) are $N_r = 365$, $N_\theta = 256$ and $N_x = 4096$, respectively. The simulation has high resolution with streamwise grid resolution of $\Delta x = 10\eta$ at $x/D = 10$, where $\eta$ is the Kolmogorov lengthscale. By $x/D = 125$, the resolution improves to $\Delta x  < 6\eta$ so that the onus on the subgrid model progressively decreases. Readers are referred to \cite{chongsiripinyo_decay_2020} for a detailed description of the numerical methodology and grid quality. 

\subsection{Spectral proper orthogonal decomposition (SPOD)}

SPOD is the frequency-domain variant of proper orthogonal decomposition. It computes monochromatic modes that are optimal in terms of the energy norm of the flow, e.g., turbulent kinetic energy (TKE) for the wake flow at hand. The SPOD modes are the eigenvectors of the cross-spectral density matrix, which is estimated using Welch's approach \citep{welch1967use}. Here, we provide a brief overview of the method. For a detailed mathematical derivation and algorithmic implementation, readers are referred to \cite{towne_spectral_2018} and \cite{schmidt2020guide}.

For a statistically stationary flow, let $\vb{q}_i =\vb{q}(t_i)$ denote the mean subtracted snapshots, where $i =1,2,\cdots n_t$ are $n_t$ number of snapshots.   For spectral estimation, the dataset is first segment  into $n_{\rm{blk}}$ overlapping blocks with $n_{\rm{fft}}$ snapshots in  each block. The neighbouring blocks overlap by $n_{\rm{ovlp}}$ snapshots with $n_{\rm{ovlp}}=n_{\rm{fft}}/2$. The $n_{\rm{blk}}$ blocks are then Fourier transformed in time and all Fourier realizations of the $l$-th frequency, $\vb{q}^{(j)}_l$, are arranged in a matrix,

\begin{equation}
\Hat{\vb{Q}}_{l}=\bqty{\Hat{\vb{q}}_{l}^{(1)}, \Hat{\vb{q}}_{l}^{(2)}, \cdots, \Hat{\vb{q}}_{l}^{(n_{\rm blk})} }.
\label{eq:fourier realizatin}
\end{equation}

The SPOD eigenvalues, $\vb*{\Lambda}_{l}$, are obtained by solving the following eigenvalue problem:
\begin{equation}
\frac{1}{n_{\rm blk}}\vb{\Hat{Q}}_{l}^{*}\vb{W} \vb{\Hat{Q}}_{l} \vb*{\Psi}_{l}=\vb*{\Psi}_{l} \vb*{\Lambda}_{l},
\label{eq:eigenvalue}
\end{equation}
where $\vb{W}$ is a positive-definite Hermitian matrix that accounts for the component-wise and numerical quadrature weights and $(\cdot)^*$ denotes the complex conjugate. SPOD modes at the $l$-th frequency are recovered as  

\begin{equation}
\vb*{\Phi}_{l} = \frac{1}{\sqrt{n_\textrm{blk}}}\vb{\Hat{Q}}_{l} \vb*{\Psi}_{l} \vb*{\Lambda}_{l}^{-1/2}.
\label{eq:SPOD_modes}
\end{equation}

The SPOD eigenvalues are denoted by $\vb*{\Lambda}_{l}=\text{diag} ( \lambda_{l}^{(1)}, \lambda_{l}^{(2)}, \cdots , \lambda_{l}^{(n_{\rm blk})} ) $. By construction,  $\lambda_{l}^{(1)} \geq \lambda_{l}^{(2)} \geq \cdots \geq \lambda_{l}^{(n_{\rm blk})}$ represent the energies of the corresponding SPOD modes that are given by the column vectors of the matrix $\vb*{\Phi}_{l}=[ \vb*{\phi}_{l}^{(1)}, \vb*{\phi}_{l}^{(2)}, \cdots , \vb*{\phi}_{l}^{(n_{\rm blk})} ]$. The SPOD mode $\vb*{\phi}_{l}^{(j)}$ represents the $j$-th dominant  coherent flow structure at the $l$-th frequency. An useful property of the SPOD modes is their orthogonality;  the weighted inner product at each frequency, $\big<\vb*{\phi}_{l}^{(i)},\vb*{\phi}_{l}^{(j)}\big>=\pqty{\vb*{\phi}_{l}^{(i)}}^*\vb{W} \vb*{\phi}_{l}^{(j)}= \delta_{ij}$.

Here, we perform SPOD on various 2D streamwise planes ranging from $x/D$ = $1$ to $120$. {Thus,  $\vb{q}$ contains the three velocity components at the discretized grid nodes on a streamwise-constant plane.} These planes are sampled at a spacing of $5D$ from $x/D = 5$ to $x/D = 100$, and five additional planes are sampled at $x/D=1, 2, 3, 110$, and $120$. The utilized time series has  $n_t = 7200$ snapshots with a uniform separation of non-dimensional time of $\Delta t U_{\infty}/D = 0.072$ between two snapshots. Owing to the periodicity in the azimuthal direction,  the flow field is first decomposed into azimuthal wavenumbers $m$, 

\begin{equation}
    q(x,r,\theta,t) = \sum\limits_{m} \hat{q}_m(x,r,t),
\end{equation}
and then  SPOD is applied on the data at each azimuthal wavenumber. The spectral estimation parameters used here are $n_{\rm{fft}}=512$ and $n_{\rm{ovlp}}=256$, resulting in $n_{\rm{blk}} = 27$ SPOD modes for each $\St$. Each block {used for the temporal FFT}  spans a time duration of $\Delta T = 36.91 U_\infty/D$.

\subsection{Reconstruction using convolution approach}
The convolution strategy proposed by \cite{nekkanti2021frequency} is employed for low-dimensional reconstruction of the flow field. This involves computing the 
expansion coefficients by convolving the SPOD modes over the data one snapshot at a time, 
    \begin{equation}
     \vb{a}_l^{(i)}(t) = \int_{\Delta T} \int_{\Omega} \qty( \vb{\phi}_l^{(i)}(x))^{*}\textbf{W}(x)\textbf{q}(x,t+\tau)w(\tau) e^{-i 2\pi f_l\tau}\dd x \dd \tau.
     \label{eq exp_conv}
    \end{equation}
Here, $w(\tau)$ is the Hamming windowing function and $\Omega$ is the spatial domain of interest. The data at time $t$ is then reconstructed as
\begin{equation}
\textbf{q}(t) \approx \sum\limits_{l}\sum\limits_{i} a^{(i)}_l (t) \vb*{\phi}^{(i)}_{l} e^{-i 2\pi f_l t}
\label{reconstruction_convolution}.
\end{equation}


\subsection{Bispectral mode decomposition}
Bispectral mode decomposition (BMD) is a technique recently proposed by \cite{schmidt2020bispectral}, to characterize the coherent structures associated with triadic interactions in statistically stationary flows. Here, we provide a brief overview of the method. The reader is referred to \citet{schmidt2020bispectral} for further details of the derivation and mathematical properties of the method. 

BMD is an extension of classical bispectral analysis to multidimensional data. The classical bispectrum is defined as the double Fourier transform of the third moment of a time signal. For a time series, $y(t)$ with zero mean, the bispectrum is 

\begin{equation}
    S_{yyy}(f_1,f_2)=\int\int R_{yyy}(\tau_1,\tau_2) e^{-i 2\pi (f_1 \tau_1 +f_2 \tau_2) }d\tau_1 d\tau_2,
\end{equation}
where $R_{yyy}(\tau_1,\tau_2)=E[y(t) y(t-\tau_1) y(t-\tau_2)]$ is the third moment of $y(t)$,  and $E[\cdot]$ is the expectation operator. The bispectrum is a signal processing tool for one-dimensional time series which only measures the quadratic phase coupling at a single spatial point. In contrast, BMD
identifies the intensity of the triadic interactions over the spatial domain of interest and extracts the corresponding spatially coherent structures.

BMD maximizes the spatial integral of the  point-wise bispectrum,
\begin{equation}
 b(f_k,f_l) = E\bqty{\int_\Omega \hat{\vb{q}}^{*}(x,f_k) \circ \hat{\vb{q}}^{*}(x,f_l) \circ \hat{\vb{q}}(x,f_k+f_l)  dx }.
\end{equation}

Here, $\hat{\vb{q}}$ is the temporal Fourier transform of $\vb{q}$ computed using the Welch approach \citep{welch1967use} and $\circ$ denotes the Hadamard (or element-wise) product.

Next, as in equation (\ref{eq:fourier realizatin}), all the Fourier realizations at $l$-th frequency are arranged into the matrix, $\hat{\vb{Q}}_{l}$. The auto-bispectral matrix is then computed as 

\begin{equation}
    \vb{B} = \frac{1}{n_{\rm{blk}}}\hat{\vb{Q}}_{k\circ l}^{H}\vb{W}\hat{\vb{Q}}_{k+l} \; , 
\end{equation}

where, $\hat{\vb{Q}}_{k\circ l}^{H} = \hat{\vb{Q}}_{k}^{*}\circ \hat{\vb{Q}}_{l}^{*}$. 

To measure the interactions between different quantities, we construct the cross-bispectral matrix
\begin{equation}
    \vb{B}_c = \frac{1}{n_{\rm{blk}}}\pqty{\hat{\vb{Q}}_{k}^{*}\circ\hat{\vb{R}}_{l}^{*}} \vb{W}\hat{\vb{S}}_{k+l}.
\end{equation}
In the present application, matrices $\vb{Q}$, $\vb{R}$, and $\vb{S}$  comprise the time series of the field variables at the azimuthal wavenumber triad, [$m_1$, $m_2$, $m_3$]. 

Owing to the non-Hermitian nature of the bispectral matrix, the optimal expansion coefficients, $\vb{a}_1$ are obtained by maximising the absolute value of the Rayleigh quotient of $\vb{B}$ (or $\vb{B}_c$)

\begin{equation}
    \vb{a}_1 = \arg\max\limits_{\|\vb{a}\|=1} \bigg |\frac{\vb{a}^*\vb{B}\vb{a}}{\vb{a}^*\vb{a}}\bigg|.
\end{equation}

The complex mode bispectrum is then obtained as  

\begin{equation}
    \lambda_1(f_k,f_l) = \bigg |\frac{\vb{a}_1^*\vb{B}\vb{a}_1} {\vb{a}_1^*\vb{a}_1}\bigg|.
\end{equation}

Finally, the leading-order bispectral modes and the cross-frequency fields are recovered as 

\begin{align}
    \vb*{\phi}_{k+l}^{(1)} &= \hat{\vb{Q}}_{k+l}\vb{a}_{1}, \quad \text{and} \\
\vb*{\phi}_{k\circ l}^{(1)} &= \hat{\vb{Q}}_{k\circ l}\vb{a}_{1},
\end{align}
respectively.  By construction, the bispectral modes and cross-frequency fields have the same set of expansion coefficients. This explicitly ensures the causal relation between the resonant frequency triad, ($f_k$, $f_l$, $f_k+f_l$), where $\hat{\vb{Q}}_{k\circ l}$ is the cause and $\hat{\vb{Q}}_{k+l}$ is the effect. The complex mode bispectrum, $\lambda_1$, measures the intensity of the triadic interaction and the bispectral mode, $\vb*{\phi}_{k+l}$, represents the structures that results from the nonlinear triadic interaction.  

Similar to SPOD, we perform BMD on various 2D streamwise planes and use the same spectral estimation parameters of $n_{\rm{fft}}=512$ and $n_{\rm{ovlp}}=256$.  Since our focus is on interactions of different azimuthal wavenumbers, the cross-BMD method,  which computes the cross-bispectral matrix $\vb{B}_c$, is applied to the wake database. The specific interactions among different $m$ and their analysis using BMD will be presented and discussed in \S \ref{sec:BMD_results}.

\section{Flow structures} \label{sec:visualizations}

\subsection{Streaky structures in the near and far wake} 

\cite{nidhan2020spectral} showed that the near and far field of the wake of a circular disk is dominated by two distinct modes residing at (a) $m=1, \St = 0.135$ and (b) $m=2, \St \rightarrow 0$. While the former mode is the vortex shedding mode in the wake of a circular disk \citep{berger1990coherent,fuchs_large-scale_1979,cannon1993observations}, the physical origin of the latter mode remains unclear. \cite{johansson_far_2006} hint that the $m=2$ mode is linked to `very' large-scale features that twist the mean flow slowly. Motivated by the findings of \cite{nidhan2020spectral} and the discussion in \cite{johansson_far_2006}, we investigate the streamwise manifestation of the azimuthal modes $m=1$ and $m=2$. The main result is that, different from the $m =1$ mode, the $m =2$ mode is associated with streamwise-aligned streaky structures.

\begin{figure}
\centering
{\includegraphics[trim={0.0cm 21.6cm 0.0cm, 0cm},clip=true,width=1\linewidth]{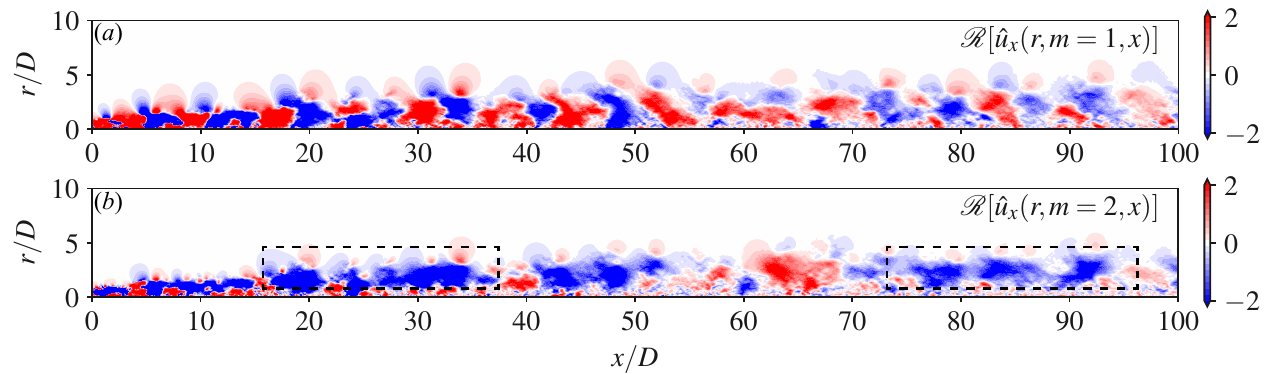}}
\caption{Azimuthally decomposed instantaneous fields of the streamwise velocity ($u_x$) for ($a$) $m = 1$ and ($b$) $m = 2$ azimuthal modes. Rectangular boxes in ($b$) show large-scale streaks in the $m=2$ field.}
\label{fig:ux_2500000_m1m2}
\end{figure}


\begin{figure}
\centering
{\includegraphics[trim={0.0cm 17.7cm, 0.0cm, 0cm},clip=true,width=1\linewidth]{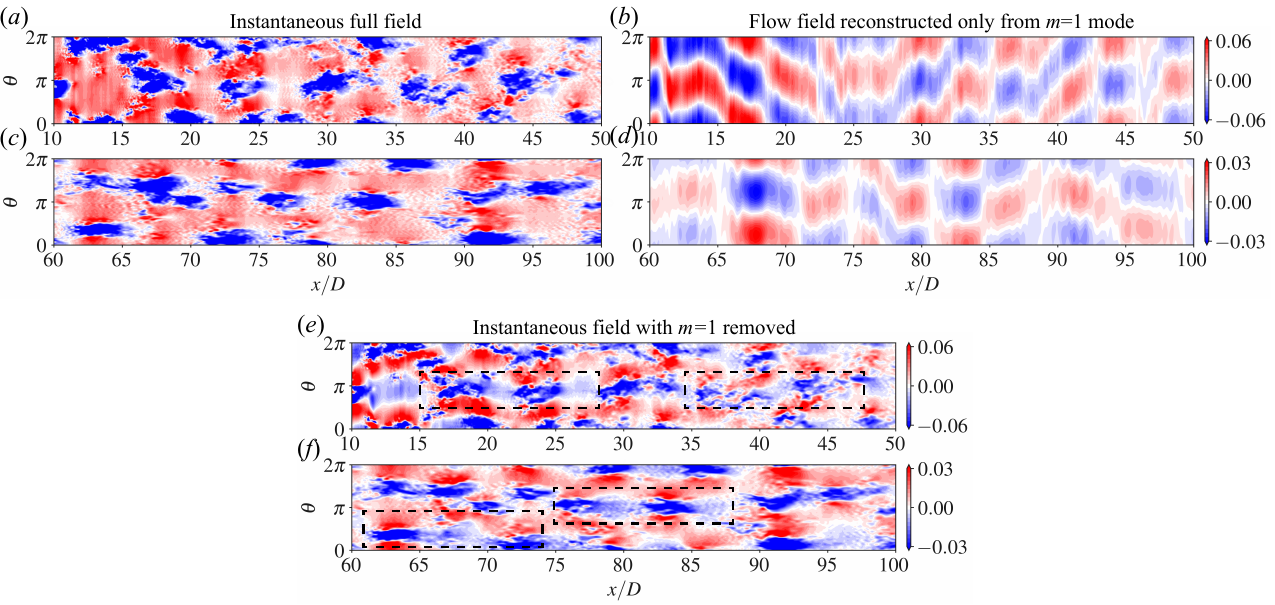}}
\caption{Instantaneous streamwise velocity on $x/D-\theta$ planes at $r/D = 1.25$ ($a$,$b$,$e$) and $r/D = 2.5$ ($c$,$d$,$f$). Panel ($a$,$c$) shows the full streamwise velocity field. Panel ($b$,$d$) shows the field reconstructed from $m=1$ contribution only. Panel ($e$,$f$) shows the field with $m=1$ contribution removed. Rectangular boxes in ($e,f$) emphasize the large-scale streaks in the flow field with the $m=1$ mode removed.}\label{fig:ux_azimuthal_inst}
\end{figure}

Figure \ref{fig:ux_2500000_m1m2} shows the azimuthal modes $m=1$ and $m=2$ of an instantaneous flow snapshot in the wake spanning downstream distance $0 < x/D < 100$, obtained using a Fourier transform in the azimuthal direction $\theta$. In the $m=1$ mode (figure \ref{fig:ux_2500000_m1m2}$a$), a wavelength of $\lambda/D = 1/\St_{\mathrm{VS}}$ (where the vortex shedding frequency, $\St_{\mathrm{VS}} \approx 0.13-0.14$) is evident throughout the domain. This observation is in agreement with previous studies \citep{johansson_proper_2002,johansson_far_2006,nidhan2020spectral} that report the existence of the vortex shedding mode at significantly large downstream locations $\sim O(100D)$ from the disk.

The spatial structure of the $m=2$ mode (figure \ref{fig:ux_2500000_m1m2}$b$) is quite different from that of the $m=1$ mode. In the $m=2$ mode visualization, distinct elongated structures are present throughout the domain. Notice in particular the structures inside the dashed rectangular boxes. The streamwise extent of these structures can be up to $\lambda_x/D \approx 25$ (see $70 < x/D < 95$), significantly larger than the wavelength of the vortex shedding mode $\lambda_x/D \approx 7$. 

Figure \ref{fig:ux_azimuthal_inst}($a$,$c$) show the instantaneous streamwise velocity field in the near-intermediate and intermediate-far wake, respectively, of the disk on a $x/D-\theta$ plane. The $x/D-\theta$ plane is constructed by unrolling the cylindrical surface at a constant $r/D$. For the near-intermediate field (figure \ref{fig:ux_azimuthal_inst}$a$),  the plane is located at $r/D = 1.25$ while for the intermediate-far field (figure \ref{fig:ux_azimuthal_inst}$c$), $r/D = 2.5$ is chosen. In both figures, vortex shedding structures are evident, spaced at an approximate wavelength of $\lambda_x/D \approx 7$. These structures and the wavelength of $\lambda_x/D \approx 7$ become even more evident when the flow field at the respective radial locations in the near-intermediate and intermediate-far wake are constructed with only the $m=1$ mode, as shown in figure \ref{fig:ux_azimuthal_inst}($b,d$). Upon removing the contribution of the vortex shedding mode that lies in $m=1$ azimuthal wavenumber \citep{johansson_proper_2002, johansson_far_2006,nidhan2020spectral}, the streaks become readily apparent in figure \ref{fig:ux_azimuthal_inst}($b$,$d$). See again the structures contained inside dashed rectangular boxes in figure \ref{fig:ux_azimuthal_inst}($b$,$d$). These observations in figure \ref{fig:ux_2500000_m1m2} and \ref{fig:ux_azimuthal_inst} are strong initial indications of the presence of large-scale streaks in turbulent wakes, similar to those recently reported in the numerical and experimental datasets of turbulent jets \citep{pickering_lift-up_2020, nogueira_large-scale_2019}. In what follows, the characteristics and robustness of these elongated structures are quantified through various statistical and spectral measures.

\begin{figure}
\centering
{\includegraphics[trim={0.0cm 3.5cm 0.0cm, 3.4cm},clip=true,width=1\linewidth]{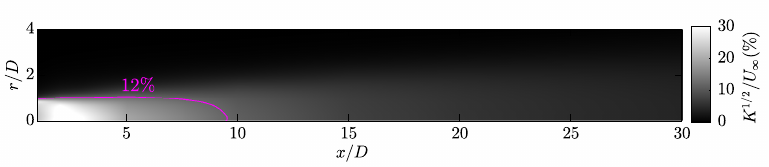}}
\caption{Turbulence intensity ($K^{1/2}/U_{\infty}$) as a function of streamwise ($x$) and radial ($r$) directions. Magenta curve demarcates the region where $K^{1/2}/U_\infty$ reduces to 0.12.}
\label{fig:Taylor_hyp}
\end{figure}

To this end, we invoke the Taylor's hypothesis, converting time $t$ at a location $x_0$ to pseudo-streamwise distance from $x_0$: $x_t = U_\mathrm{conv} t$, where $U_{\mathrm{conv}} = U_{\infty} - U_d$ is the convective velocity. Since the defect velocity in the wake $U_d \ll U_\infty$ (where $U_\infty$ is the free-stream velocity),  $U_{\mathrm{conv}}$ is approximated by $ U_{\infty}$.  Taylor's hypothesis requires velocity fluctuation  to be sufficiently small compared to $U_{\mathrm{conv}}$. Figure \ref{fig:Taylor_hyp} shows that this requirement is met since turbulence intensity ($K^{1/2}/U_\infty$)  drops below $12\%$ beyond $x/D = 10$ for all radial locations and below $ 4 \%$ at $x/D =  30$.
Previous work on turbulent wakes has shown Taylor's hypothesis to be valid in the wake when $K^{1/2}/U_{\infty}$ drops below $\sim 10\%$ \citep{antonia1998approach, kang2002universality, dairay2015non, obligado2016nonequilibrium}. Invoking Taylor's hypothesis, $\St \approx U_\infty k_x$. In this study  we will particularly focus on $\St \rightarrow 0$ (obtained from either temporal spectra or SPOD at different $x/D$ locations) which, by Taylor's hypothesis, is equivalent to large-scale streaks with $k_x \rightarrow 0$ ($k_x$ being the streamwise wavenumber) at those locations. 

\begin{figure}
\centering
{\includegraphics[trim={0.0cm 1.6cm 0.0cm 2.1cm},clip,width=1.0\textwidth]{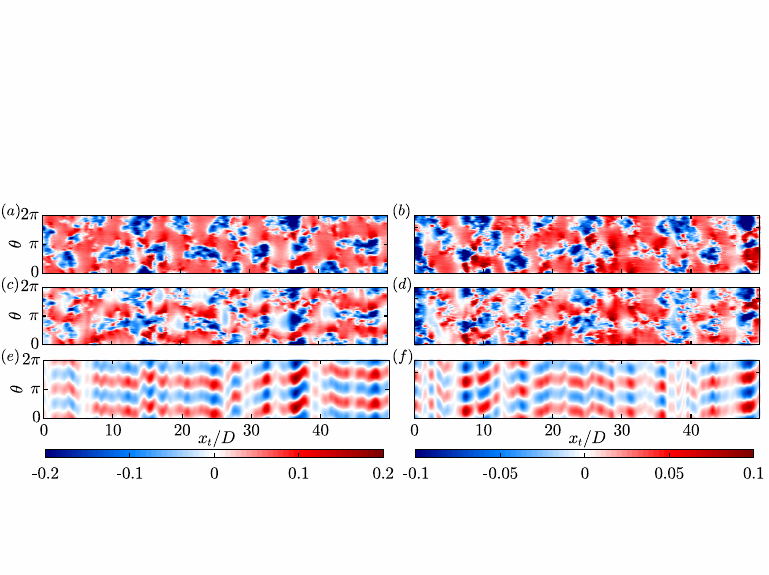}}
\caption{Streamwise velocity fluctuations $u'_x$ on a $x_t/D$--$\theta$ plane at $r/D \approx 0.8$: ($a,c,e$) $x_0/D = 10$; ($b$,$d$,$f$) $x_0/D = 20$. Top row ($a$,$b$) include all azimuthal components, middle row ($c$,$d$) excludes  $m=1$, and the bottom row ($e$,$f$)  includes solely $m=2$.}
\label{fig:Azim_x_5_10_15}
\end{figure}


In figure \ref{fig:Azim_x_5_10_15}, the near-to-intermediate wake behavior  is shown at  $x_0/D = 10$ (left column) and $20$ (right column) through plots of streamwise velocity fluctuation ($u'_x$)  at $r/D = 0.8$ in the $x_t/D - \theta$ plane. Time $t$ is transformed to the $x_t$ coordinate by application of  Taylor's hypothesis, which  is valid beyond $x/D \ge 9$ in the wake, as was shown through figure \ref{fig:Taylor_hyp} and the associated discussion. 

Due to the strong signature of the vortex shedding mode in the near to intermediate wake, alternate patches of inclined positive and negative fluctuations separated by $\lambda_D \approx 1/\St_{\mathrm{VS}}$ dominate the visualization in figure \ref{fig:Azim_x_5_10_15}($a,b$). It is known \textit{a priori} that these structures are contained in the $m=1$ azimuthal mode. In order to assess space-time coherence other than the vortex shedding mode,  the streamwise fluctuations are replotted   in figure \ref{fig:Azim_x_5_10_15}($c$,$d$) after removing the  $m=1$ contribution. Once the $m=1$ contribution is removed, streamwise streaks become evident at both locations. Furthermore, one can also observe that these streaks appear to be primarily contained in the azimuthal mode $m=2$, i.e, there are two structures over the azimuthal length of $2\pi$. Only the $m=2$ component is shown in figure \ref{fig:Azim_x_5_10_15} ($e$,$f$)  and, on doing so, the elongated streamwise streaks come into sharper focus. Building upon figure \ref{fig:ux_2500000_m1m2}, figure \ref{fig:Azim_x_5_10_15} ($c$,$d$,$e$,$f$) lend support to the spatiotemporal robustness of these large-scale streaks in the turbulent wake of circular disk.

\begin{figure}
\centering
{\includegraphics[trim={0.0cm 3.2cm 0.0cm 2.0cm},clip,width=1.0\textwidth]{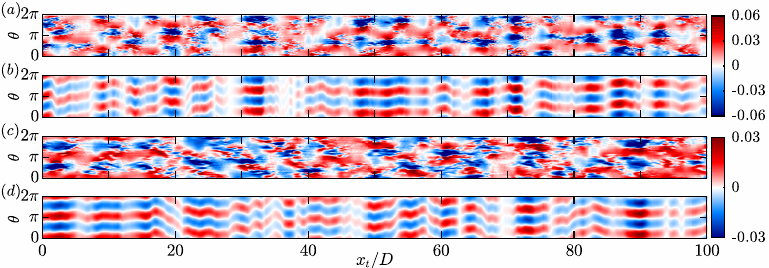}}
    \caption{The $u'_x$ field on a $x_t/D$--$\theta$ plane at $r/D \approx 2.0$: ($a$,$b$) $x_0/D = 40$; ($c$,$d$) $x_0/D = 80$. In ($a$,$c$)  $m=1$ is removed and in ($b$,$d$) only  $m=2$ is shown.}
\label{fig:Azim_x_40_60_80}
\end{figure}


The downstream distance ($0<x/D<120$)  spanned in the simulations of \cite{chongsiripinyo_decay_2020} is very large, thus enabling   the far field to be probed too for the presence (or absence) of the streamwise streaks. Figure \ref{fig:Azim_x_40_60_80}($a$,$c$) show the $x_t/D-\theta$ plots of $u'_x$, with contribution of $m=1$ mode excluded, in regions starting at $x_0/D = 40$ and $80$. The  $x_t/D-\theta$ planes are located at $r/D =2$ for these far-wake locations since the wake width grows  with $x$. Interested readers can refer to figures 5, 15, 16 of \cite{chongsiripinyo_decay_2020} and figure 5 of \cite{nidhan2020spectral} for an in-depth discussion about the mean and turbulence statistics. Similar to the near wake plot in figure \ref{fig:Azim_x_5_10_15}($c$,$d$), large-scale streaks elongated in the streamwise direction are found to extend into  the far wake as well. Once again, isolating  the $m=2$ component  (figure \ref{fig:Azim_x_40_60_80}$b$ and $d$) highlights the streaks. Collectively, figures \ref{fig:Azim_x_5_10_15}, and \ref{fig:Azim_x_40_60_80} demonstrate that the  streaks span the entire wake length and that the $m = 2$ mode drives these streaks.

\begin{figure}
\centering
{\includegraphics[trim={0.0cm 1.2cm 0.0cm 1.5cm},clip,width=1.0\textwidth]{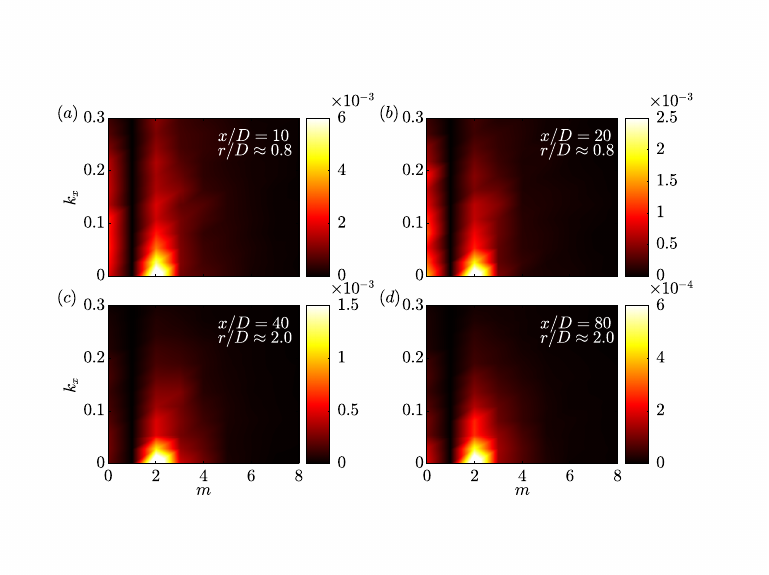}}
\caption{2D power spectral density of the streamwise velocity  fluctuation in ($k_x$, $m$) space: ($a$) $x/D = 10$, $r/D \approx 0.8$; ($b$) $x/D = 20$, $r/D \approx 0.8$; ($c$) $x/D = 40$, $r/D \approx 2.0$; ($d$) $x/D = 80$, $r/D \approx 2.0$. The $m=1$ wavenumber is removed so as to de-emphasize the vortex-shedding structure.}
\label{fig:spectra_kx_m_x_5_10_40_80}
\end{figure}

Figure \ref{fig:spectra_kx_m_x_5_10_40_80} shows  two-dimensional (2D) spectra in the streamwise wavelength ($k_x$) -- azimuthal mode ($m$) space at four representative streamwise locations $x/D = 10, 20, 40,$ and $80$. Here, $k_x$ is  the wavenumber of the pseudo-streamwise direction $x_t$.  The $m=1$ contribution is removed \textit{a priori} to emphasize the large-scale streaks. At all these four locations, these streaky structures correspond to $k_x \rightarrow 0$ and are found to reside in the $m=2$ azimuthal mode. Note that the $k_x = 0, \St=0$ should be interpreted as $k_x, \St \rightarrow 0$ as the length of the time series is not sufficient to resolve the large time-scale of streaks. In appendix \ref{sec:appA}, we vary the spectral estimation parameter $n_{\rm fft}$ to resolve the lower frequencies and identify the frequency associated with streaks in this limit to be around $\St \approx 0.006$. 


\subsection{Evidence of lift-up mechanism in the wake}

\begin{figure}
\centering
{\includegraphics[trim={0.0cm 14.0cm 0.0cm, 0cm},clip=true,width=1\linewidth]{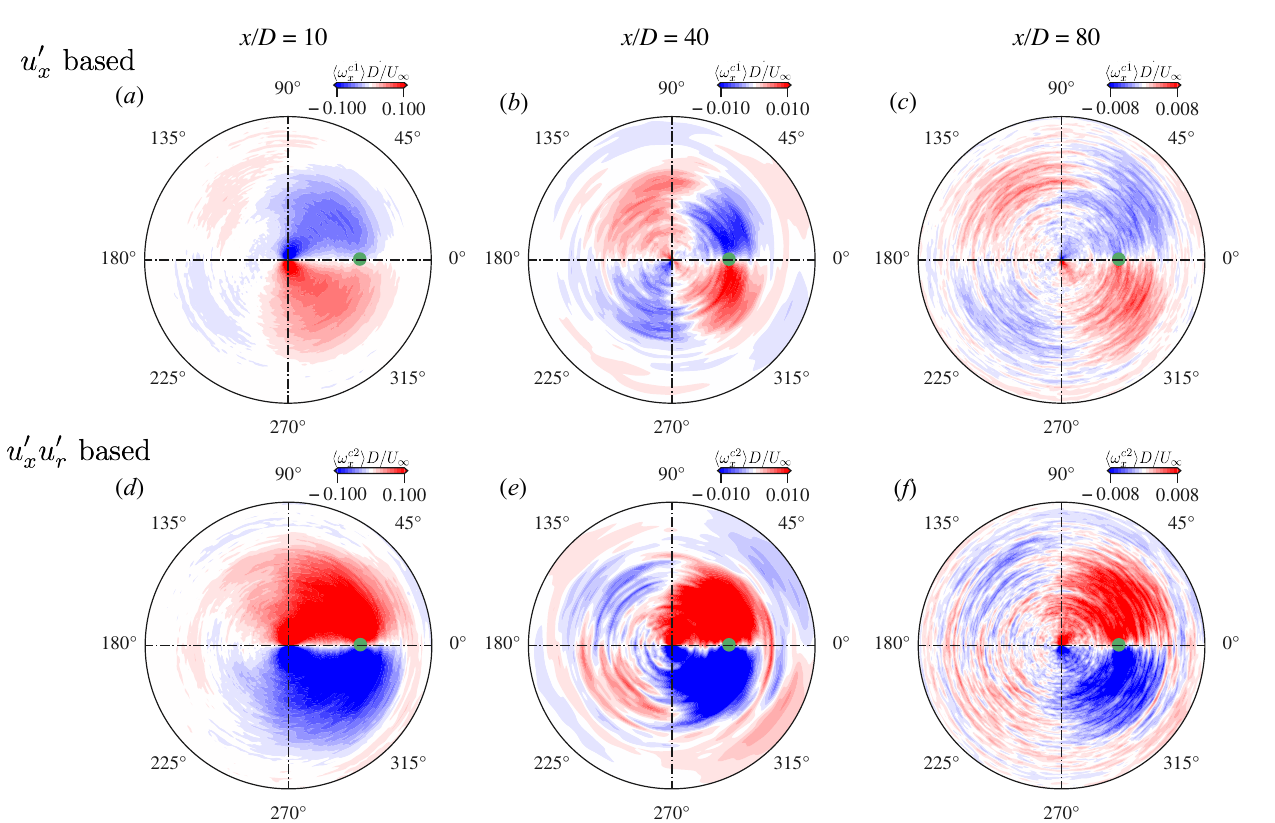}}
\caption{Conditionally-averaged streamwise vorticity at $x/D = 10$ (left column), $40$ (middle column)  and $80$ (right column).  Top row ($a,b,c$) shows $\langle \omega^{c1}_x \rangle$, conditioned using  $u'_x$ and bottom row ($d,e,f$) shows $\langle \omega^{c2}_x \rangle$, conditioned  using $u'_xu'_r$. Green dot shows the location of conditioning point -- $(a,d)$ $r/D \approx 0.8, \theta \approx 0^{\circ}$ and $(b,c,e,f)$ $r/D \approx 2, \theta \approx 0^{\circ}$. Radial domain extends until  $r/D = 2$ at  $x/D =10$ and until  $r/D = 5$ at $x/D=40,80$. Positive vorticity (red) is out of the plane and negative vorticity (blue) is into the plane.}
\label{fig:conditional_avg_omegax}
\end{figure}

Previous work in turbulent free shear flows and wall-bounded flows 
often attribute the formation of  streaks in the velocity to the  lift-up mechanism \citep{ellingsen_stability_1975,landahl1975wave}. The lift-up mechanism, by sweeping  fluid from high-speed regions to  low-speed regions and vice versa,  leads to the formation of high-speed and low-speed streaks, respectively. \cite{brandt2014lift} provides a comprehensive review of the lift-up mechanism and its crucial role in various fundamental phenomena, e.g., subcritical transition in shear flows, self-sustaining cycle in wall bounded flows, and disturbance growth in complex flows. 

To investigate the presence of the lift-up mechanism in the wake, we plot  conditional averages of  streamwise vorticity fluctuations ($\omega_x$)  at three representative locations in the flow -- the planes, $x/D = 10, 40$ and $80$ in figure \ref{fig:conditional_avg_omegax}. The top row  shows a conditional average,  $ \langle \omega^{c1}_x \rangle$,   designed to extract the structure of the streamwise vorticity on a constant-$x$ plane  during times of large streamwise velocity fluctuations. Specifically, the condition is that  
\begin{equation}
    u'_x \ge c(u'_x)^{\mathrm{rms}},
    \label{eq:ux_condition}
\end{equation}
 at a specified point P on that plane, and $ \langle \omega^{c1}_x \rangle$ is the temporal average of all $\omega_x (t)$  that satisfies this condition. The conditional point P (shown as green dots in figure \ref{fig:conditional_avg_omegax}) is chosen to  lie at $\theta = 0$ and radial locations of $r/D = 0.8$ and $2$ for $x/D = 10$ and $x/D=40,80$, respectively. The selection of different radial locations at $x/D = 10$ and $x/D=40,80$ is based on the approximate values of mean wake half-widths at the respective $x/D$ locations \citep{chongsiripinyo_decay_2020}. Owing to  rotational invariance of  statistics for an axisymmetric wake, the condition is applied to a new  point $\rm{P}_1$  at  the same $r/D$ but a different value of $\theta$ and the new  $ \langle \omega^{c1}_x \rangle$ field, after a rotation to bring $\rm{P}_1$ to P, is included in the conditional average. Since  $N_\theta = 256$ points is used for discretization, the ensemble used for the conditional average is significantly expanded by exploiting  rotational invariance of statistics. The bottom row of figure  \ref{fig:conditional_avg_omegax} shows $ \langle \omega^{c2}_x \rangle$ computed using a different condition at point P, 
\begin{equation}
    -u'_xu'_r \ge c(-u'_xu'_r)^{\mathrm{rms}} \; .
    \label{eq:uxur_condition}
\end{equation}
 This condition is designed to identify the structure of streamwise vorticity at times of significant Reynolds shear stress at point P. The results exhibit moderate sensitivity to $c \in (0, 1]$ as reported in appendix \ref{sec:appB}. Hence, $c$ is set to $0.5$ as a compromise between identification of  intense events and retention of sufficient snapshots for conditional averaging. $(u'_x)^{\mathrm{rms}}$ and $(-u'_x u'_r)^{\mathrm{rms}}$ are the r.m.s. values of the streamwise velocity fluctuations and the r.m.s. values of the streamwise-radial fluctuations correlation at the conditioning points, respectively.

The conditionally averaged field based on equation (\ref{eq:ux_condition}) captures the structure of the streamwise vorticity field during events of intense positive  $u'_x$. In figure \ref{fig:conditional_avg_omegax}($a$), two rolls of streamwise vorticity are observed in the conditionally averaged field: negative on the top and positive at the bottom of the conditioning point, respectively. These streamwise vortex rolls push the high-speed fluid in the outer wake to the low-speed region in the inner wake around the conditioning points (green dot), leading to $u'_x > 0$. When the averaging procedure is conditioned on negative streamwise velocity fluctuations, i.e., $u'_x \leq -c(u'_x)^{\mathrm{rms}}$, the signs of the vortex rolls in figure \ref{fig:conditional_avg_omegax}  are interchanged, as expected (figure not shown). At $x/D = 40$ and $80$ (figure \ref{fig:conditional_avg_omegax}$b,c$), two additional vortical structures are observed in the $\theta = [90^{\circ}, 270^{\circ}]$ region. However, around the conditioning point the spatial organization of vorticity remains qualitatively similar. The size of these vortex rolls increase with $x/D$, consistent with the radial spread of wake.


{The conditionally averaged field based on equation (\ref{eq:uxur_condition}) captures the vorticity field corresponding to intense positive $-u'_xu'_r$ values. In a turbulent wake, $-u'_xu'_r$ is predominantly positive such that the dominant production term in the wake, $P_{xr} = \langle -u'_xu'_r \rangle \partial U /\partial r > 0$, acts to  transfer energy from the mean flow to turbulence. Now turning to equation (\ref{eq:uxur_condition}), positive $-u'_xu'_r$ can result from two scenarios: (a) ($u'_r > 0$, $u'_x < 0$), i.e., `ejection' of low-speed fluid from the inner wake to the outer wake and (b) ($u'_r < 0$, $u'_x > 0$), i.e., `sweep' of high-speed fluid from outer wake to the inner wake. Both of the above-mentioned scenarios are consistent with the lift-up mechanism. If ejection and sweep events were equally probable, $\langle \omega^{c2}_x \rangle \approx 0$ due to the opposite spatial distribution of vortices during ejection and sweep events. However, $\langle \omega_x^{c2}\rangle$ obtained from $-u'_xu'_r$ based conditioning (bottom row of figure \ref{fig:conditional_avg_omegax}) shows that the positive and negative vortices are spatially organized such that the flow induced by these vortices at the conditioning point is outward ($u'_r > 0$), pushing low-speed fluid from the inner wake to the outer wake. In short, $\langle \omega^{c2}_{x} \rangle$ fields in the bottom row of figure \ref{fig:conditional_avg_omegax} correspond to the ejection events at the conditioning point. A similar spatial organization of $\langle \omega^{c2}_x \rangle$ is observed across the wake cross-section when the radial location of the conditioning point is varied (plots not shown for brevity). This observation establishes that ejections are the dominant contributors to intense positive $-u'_xu'_r$, as opposed to sweep events, and therefore, ejections are more  instrumental in the energy transfer from mean to turbulence. Previous studies \citep{wallace2016quadrant,kline1967structure,corino1969visual} of the turbulent boundary layer have also reported  that ejection events are the primary contributors to Reynolds shear stress.

Figure \ref{fig:conditional_avg_omegax} has two important implications. First, the top row demonstrates strong correlation between intense $u'_x$ fluctuations and distinct streamwise vortical structures, indicating that the lift-up mechanism is active in the turbulent wake, both in the near field as well as the far field. Second, the conditionally averaged fields obtained using $u'_xu'_r$ inform us that the lift-up mechanism corresponding to the ejection of low-speed fluid from the inner wake to the outer wake is more dominant than the sweep of high-speed fluid from the outer wake to the inner wake. To the best of authors' knowledge, both these observations constitute the first numerical evidence in the near and far field of a canonical bluff-body turbulent wake of  (a) the lift-up mechanism and (b) the dominance of ejection events .

\section{SPOD analysis of streaks in the wake} \label{sec:spod_analysis}

\S\ref{sec:visualizations} reveals the presence of large-scale streaks   and also that the lift-up mechanism is active in the wake. Furthermore, the $m=2$ azimuthal wavenumber visually appears to be the dominant streak-containing mode. In this section, SPOD {is employed} to quantify the energetics and educe the dominant structures of the dominant features at $\St \rightarrow 0$ in the wake. We particularly focus on the $m=2$ mode, providing further evidence that these modes exhibit properties of streaks and are formed due to the lift-up mechanism. {Direct comparison with the VS mode ($m =1$, $St = 0.135$) is provided as appropriate to differentiate the role of streaks from that of the VS mode.} 



\subsection{Energetics of streaky structures using SPOD analysis}\label{spod:streaks}

\begin{figure}
\centering
{\includegraphics[trim={0.0cm 1.6cm 0.0cm 2.cm},clip,width=1.0\textwidth]{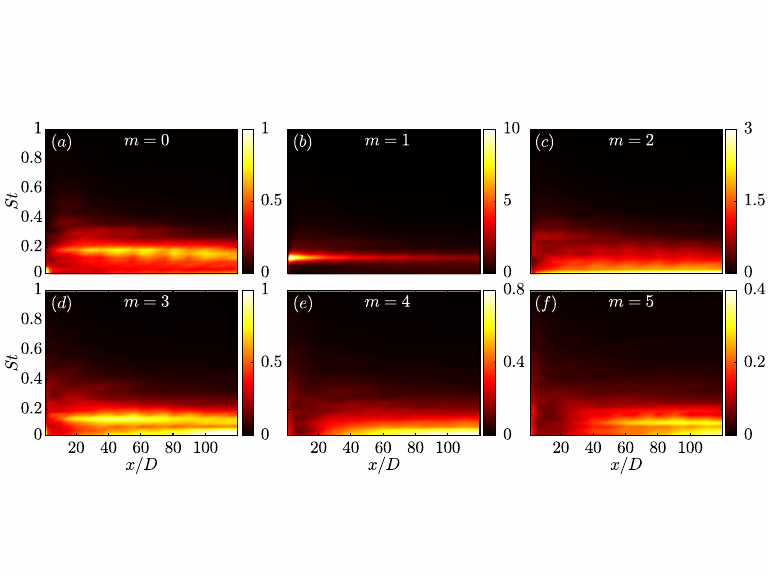}}
\caption{Percentage of energy contained in the leading SPOD modes, $\lambda^{(1)}$, as a function of frequency $\St$ and streamwise location $x/D$ at different azimuthal wavenumbers:  ($a$) $m = 0$; ($b$) $m = 1$; ($c$) $m = 2$; ($d$) $m = 3$; ($e$) $m = 4$; ($f$) $m = 5$. The leading SPOD eigenvalue, $\lambda^{(1)}(St,x/D)$, is normalized with the total TKE $E_k^{T}(x/D)$ at the corresponding streamwise location to calculate the percentage contribution. }
\label{fig:SPOD_eigenval_contour}
\end{figure}

SPOD is performed at different streamwise locations ($x/D$) in $1 \le x/D \le 120$.  By definition, the leading    SPOD modes at a given $x/D$   represents the most energetic coherent structures at the associated frequency ($St$) and azimuthal wave number ($m$). Figure \ref{fig:SPOD_eigenval_contour} shows the percentage of energy in the leading SPOD modes ($\lambda^{(1)}$) as a function of frequency and streamwise distance for the first six azimuthal wavenumbers $m = 0$ to $m=5$. The percentage of energy at each streamwise location is obtained by normalizing the leading eigenvalue with the total turbulent kinetic energy, $E^{T}_k(x/D)$ at the corresponding location. Overall, the most significant contributors to the TKE are the vortex shedding mode ($m=1$, $\St=0.135$) and 
the mode corresponding to streaks ($m=2$, $\St \rightarrow 0$), as reported in \cite{nidhan2020spectral}. The leading vortex shedding SPOD mode contains about 10\% energy in the near wake region ($5 \lesssim x/D \lesssim 15$) and decreases thereafter. The leading SPOD mode corresponding to the streaks in the $m=2$ mode contains approximately 3\% energy from $x/D=10$ onward. The axisymmetric component ($m=0$) exhibits a peak at $\St \rightarrow 0.054$ at $x/D=1$ and a much smaller peak at $\St \approx 0.19$ between $10 \lesssim x \lesssim 120$. The former is associated with the pumping of the recirculation bubble \citep{berger1990coherent}, whereas the latter was observed in previous studies (see figure 12 in \cite{berger1990coherent} and figure 7 in \cite{fuchs1979large}) but was not investigated further. 
{The $m = 0$ mode is not the focus of this study}.   

The energy, $\lambda^{(1)}$,  contained in the higher azimuthal wavenumbers ($m=3-5$) is shown in figures \ref{fig:SPOD_eigenval_contour}($d$-$f$), respectively. Although $\lambda^{(1)}$  is smaller than at  $m=1$ or $m=2$, the higher modes also exhibit temporal structure. The $m=3$ component shows energy concentration at the vortex shedding frequency $\St=0.135$ for $15 \lesssim x/D \lesssim 70$ and at $\St \rightarrow 0$ for $x/D \gtrsim 50$. For $m=4$, energy is concentrated near $\St \rightarrow 0$ for $x/D \gtrsim 60$. For the $m=5$ component, traces of the vortex shedding mode and streaks ($\St \rightarrow 0$) are observed at the streamwise locations $x/D \gtrsim 60$ and $x/D \gtrsim 80$, respectively. Figure \ref{fig:SPOD_eigenval_contour} indicates that the peaks at the vortex shedding frequency are present only at the odd azimuthal wavenumbers ($m=1,3,5$), whereas the peaks corresponding to the large-scale streaks, i.e., $\St \rightarrow 0$, can be found at both the odd and even $m$. It is also interesting to note that, for higher $m$, both the vortex shedding modes and streaks do not appear until larger values of $x/D$. This suggests nonlinear interactions among different frequencies and azimuthal wavenumbers as the wake evolves, as will be elaborated in \S\ref{sec:nonlinear_interaction}. 

\begin{figure}
\centering
{\includegraphics[trim={0.0cm 3.4cm 0.0cm 1.55cm},clip,width=1.0\textwidth]{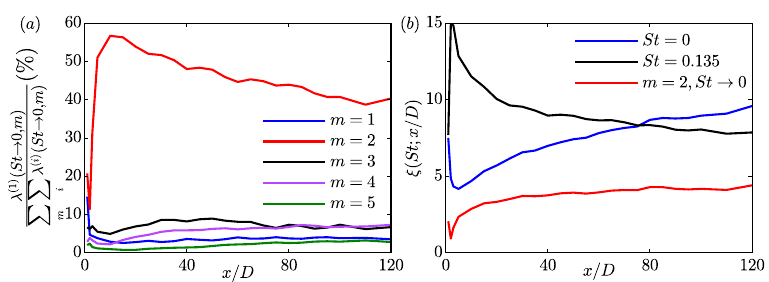}}
\caption{($a$) Contribution of the leading SPOD modes $\lambda^{(1)}$ at different azimuthal wavenumbers to $\St \rightarrow 0$. ($b$) Percentage energy contained in the frequencies $\St=0.135$ and $\St\rightarrow0$.}
\label{fig:energy_percentage}
\end{figure}

{The $\St \rightarrow 0$ streaks are dominated by the $m =2$ azimuthal wavenumber as demonstrated } by figure \ref{fig:energy_percentage}($a$), which shows the contribution of different $m$ at $\St \rightarrow 0$. The leading eigenvalues of each azimuthal wavenumber is normalized by the total energy at $\St \rightarrow 0$, i.e., $ \sum_m \sum_i  \lambda^{(i)} (St\rightarrow 0,m)$. Streaks are azimuthally non-uniform structures and are not present in the axisymmetric  $m=0$ component. Hence, we focus on  $m \ge 1$. The azimuthal wavenumber $m=2$ is energetically dominant at the $\St \rightarrow 0$ frequency, containing about $40-50\%$ of the total energy at $\St \rightarrow 0$. The sub-optimal wavenumber  is $m=3$ for $5\le x \le80$ and switches between $m=3$ and $m=4$, thereafter. However, the difference in energy between the $m=2$ and $m=3$ wavenumbers is always large,  $>$$30\%$. This dominance of the $m=2$ wavenumber at $\St \rightarrow 0$ is also consistent with the visualizations of the streamwise velocity fluctuations in figure \ref{fig:Azim_x_5_10_15} and figure \ref{fig:Azim_x_40_60_80} where one can even see the presence of $m=2$ with the naked eye.

The energetic contribution of the  two dominant frequencies, $\St=0.135$ (vortex shedding) and $\St \rightarrow 0$ (streaks),   is compared  by computing the percentage of total energy (across all $m$ and $i$) at a given frequency $\St$ as 
\begin{equation}
    \xi(\St;x/D)=\frac{\sum\limits_m \sum\limits_i \lambda^{(i)}(m,\St;x/D)}{E^{T}_k(x/D)} \times 100. 
\end{equation}
Figure \ref{fig:energy_percentage}($b$) shows that the vortex shedding frequency is more dominant in the region $x\le 70$, whereas the zeroth frequency dominates for $x\ge70$. This implies that although the streaks are present throughout, they are energetically more prominent in the far wake region.  It is interesting to note that beyond $x/D \approx 65$, the defect velocity decay rate changes from $x^{-1}$ to $x^{-2/3}$ in the wake \citep{chongsiripinyo_decay_2020}. For comparison, the most dominant component at the zeroth frequency, i.e.,  $m=2$ is also shown in figure \ref{fig:energy_percentage}($b$), which exhibits a similar trend of increasing prominence in the downstream direction. 

\begin{figure}
\centering
{\includegraphics[trim={0.0cm 1.25cm 0.0cm 1.25cm},clip,width=1.0\textwidth]{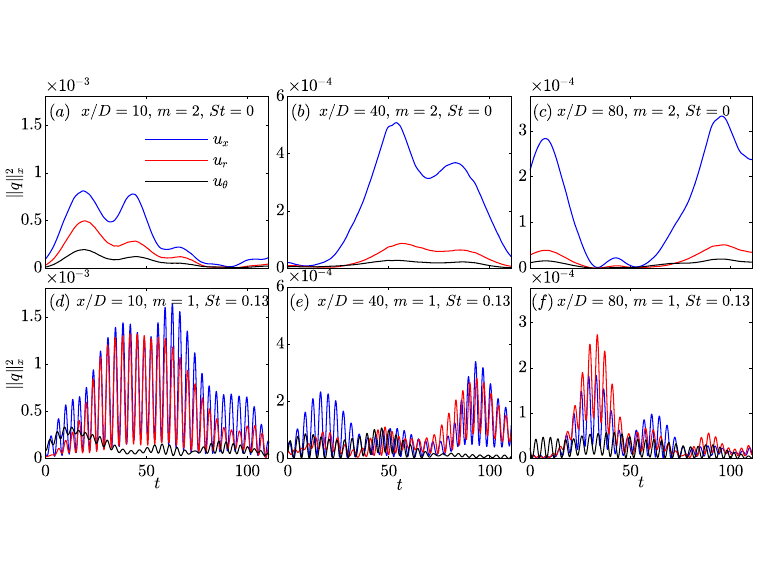}}
\caption{Component-wise instantaneous energy reconstructed from the  $m =2$  streaky structures in the top row ($a$-$c$) and vortex shedding structures ($d$-$f$) in the bottom row. The plane integral of the energy is shown at various streamwise locations:  ($a,d$) $x/D = 10$, ($b,e$) $x/D = 40$ and ($c,f$) $x/D = 80$. Flow reconstruction uses the leading five SPOD modes. The oscillation frequency of the reconstructed energy in ($d$-$f$) is twice that of the VS frequency.} 
\label{fig:reconstruction_20_40_80}
\end{figure}

Having diagnosed the streaky structure ($m=2, \St\rightarrow 0$) and the vortex shedding structure ($m=1, \St=0.135$)  using SPOD, we shift focus to their imprint on the flow in physical space by reconstructing the flow field using their leading five SPOD modes.  
The $m=2$ wavenumber is selected because it is energetically dominant at $\St \rightarrow 0$. The reconstruction is performed using the convolution strategy described in section \S\ref{sec:methodology_database}. Figure \ref{fig:reconstruction_20_40_80}($a$,$b$,$c$) shows the instantaneous energy in the three fluctuation components, $u'_x$, $u'_r$, and $u'_{\theta}$ at  $x/D =10$, 40, and 80, after reconstruction with the streaky-structure SPOD modes. The instantaneous energy is depicted within the time interval $t \in [0, 110]$, which corresponds to the first five blocks used for SPOD, and is representative of the entire reconstructed flow fields. The energy of $u'_x$ is significantly higher than that of  $u'_r$ and $u'_\theta$. The dominance of the streamwise component over its radial and azimuthal counterparts is one salient feature of streaks. The finding of $u'_x$ dominance is in agreement with those of \citet[see figure 11]{boronin2013non} and \citet[see figure 10]{pickering_lift-up_2020} in the context of round jets, indicating similarity in streak formation between jets and wakes.  The lift-up mechanism, in which the streamwise vortices lift up and push down the low-speed and high-speed fluid, respectively,  is responsible for elongated streaks accompanied by an amplification of $u'_x$. 
The analog of panels in figure \ref{fig:reconstruction_20_40_80}($a$-$c$) is shown for the VS mode in panels of figure \ref{fig:reconstruction_20_40_80}($d$-$f$). 
Here,  $u'_x$ and $u'_r$ components have comparable energy, showing a fundamental  difference between the vortex shedding mode and the streaky-structures mode as to how each mode contributes to velocity fluctuations in the wake. Furthermore, the instantaneous energy of the flow field reconstructed from the VS mode has a much smaller time scale in comparison to that of the large-scale streaks. 

\begin{figure}
\centering
{\includegraphics[trim={0.0cm 0.95cm 0.0cm 0.95cm},clip,width=1.0\textwidth]{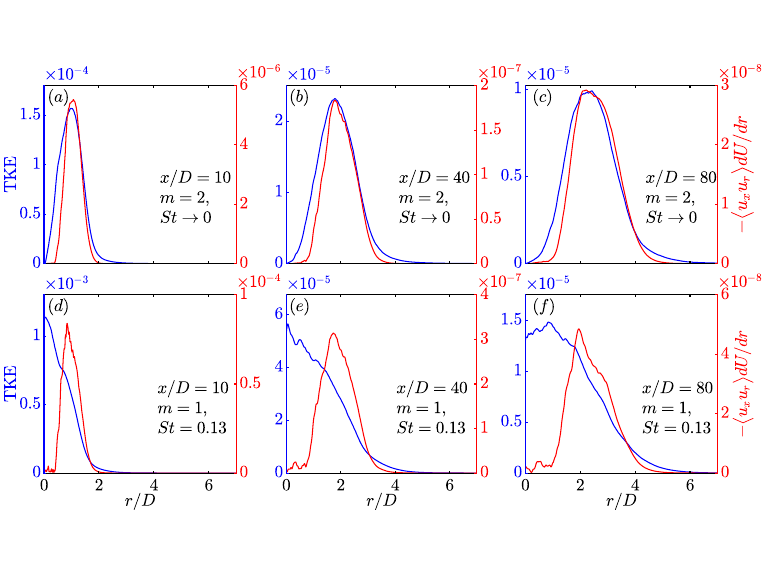}}
\caption{TKE and  production due to shear at different downstream locations for the flow fields reconstructed from the  streaky-structures mode in the top row ($a$-$c$) and VS mode in the bottom row ($d$-$f$). Left column ($a,d$) corresponds to $x/D = 10$,  middle column ($b,e$) to $x/D = 40$ and right column ($c,f$) to $x/D = 80$. Flow reconstruction uses the leading five SPOD modes.} 
\label{fig:TKE_REY_10_40_80}
\end{figure}

Figure \ref{fig:TKE_REY_10_40_80} shows the TKE ($K= \big< u'_iu'_i\big>/2$) and its shear production ($P_{xr}= -\big< u'_xu'_r\big>\partial U/\partial r$) corresponding to the large-scale streaks and vortex shedding structures at $x/D=10$, $40$ and $80$. As in figure \ref{fig:reconstruction_20_40_80}, TKE and $P_{xr}$ are computed from the leading five SPOD modes. Figure \ref{fig:TKE_REY_10_40_80}($a$-$c$) show that the TKE and the production peak at the similar radial location for $m=2$, $St\rightarrow0$. This is not the case for $m=1$, $\St=0.135$ (figure \ref{fig:TKE_REY_10_40_80}$b$-$f$) where the peak TKE occurs close to the centerline while the production peaks away from the centerline. This difference in the locations of peak $K$ and $P_{xr}$ indicates that turbulent transport plays an important role in distributing the TKE in the vortex shedding mode, similar to its importance in the full TKE budget of an axisymmetric wake \citep{uberoi1970turbulent}. The difference in radial locations of peak $K$ and $P_{xr}$, as demonstrated in figure \ref{fig:TKE_REY_10_40_80}, is another crucial distinction between the large-scale streaky mode  and the vortex shedding mode. The presence of streaks is associated with high TKE around the region of high production/mean shear  indicating their important role in the energy transfer from mean to fluctuation velocity in the turbulent wake, similar to {other shear flows} \citep{gualtieri2002scaling,brandt2007numerical,jimenez-gonzalez_transient_2017}.
As a result, the TKE achieves its global maximum around the same location as that of $P_{xr}$ for streaks.

\subsection{Lift-up mechanism through the lens of SPOD analysis} \label{spod:liftup}

\begin{figure}
\centering
{\includegraphics[trim={0.0cm 2.2cm 0.0cm 2.7cm},clip,width=1.0\textwidth]{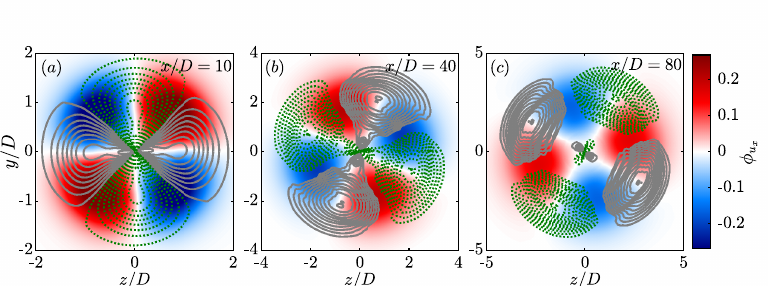}}
\caption{Leading SPOD mode of $m=2$  and $St\rightarrow0$: ($a$) $x/D=10$; ($b$) $x/D=40$; ($c$) $x/D=80$. The false colors represent the real part of the streamwise velocity component $\mathrm{Re}\bqty{\phi^{(1)}_{u_x}(St\rightarrow 0)}$ and the grey and green-dotted lines correspond to positive and negative streamwise vorticity $\mathrm{Re}\bqty{\phi^{(1)}_{\omega_x}(St\rightarrow 0)}$, respectively. }
\label{fig:spod_eigenmode_20_40_80}
\end{figure}

 \S \ref{spod:streaks} demonstrated that the structures associated with $\St \rightarrow 0$ exhibit the characteristics of streaks and their significance  increases from the near to far wake, with the azimuthal wavenumber $m=2$ being the most significant, energetically, to the streaks. Figure \ref{fig:spod_eigenmode_20_40_80} shows the leading SPOD mode of the streamwise velocity fluctuation ($u'_x$) corresponding to $m=2$ and $\St \rightarrow 0$ at three streamwise locations $x/D = 10$, $40$, and $80$. Overlaid on the $u'_x$ contour   is the streamwise vorticity ($\omega'_x$) of the corresponding mode.  Both  $u'_x$ and $\omega'_x$ are characterized by four lobes of alternate sign, the size of which increase monotonically with $x/D$.  Importantly,  the set of $\omega'_x$ lobes is shifted with  respect to the $u'_x$ lobes  by a clockwise rotation of approximately $45^{\circ}$. As a result of the shift, the maximum of $u'_x$ in the mode appears at the location where the vortices  bring in high-speed fluid from the outer to the inner wake, and vice versa. This observation further confirms the presence of lift-up mechanism in the wake.

\begin{figure}
\centering
{\includegraphics[trim={0.0cm 19.5cm 0.0cm 0cm},clip,width=1.0\textwidth]{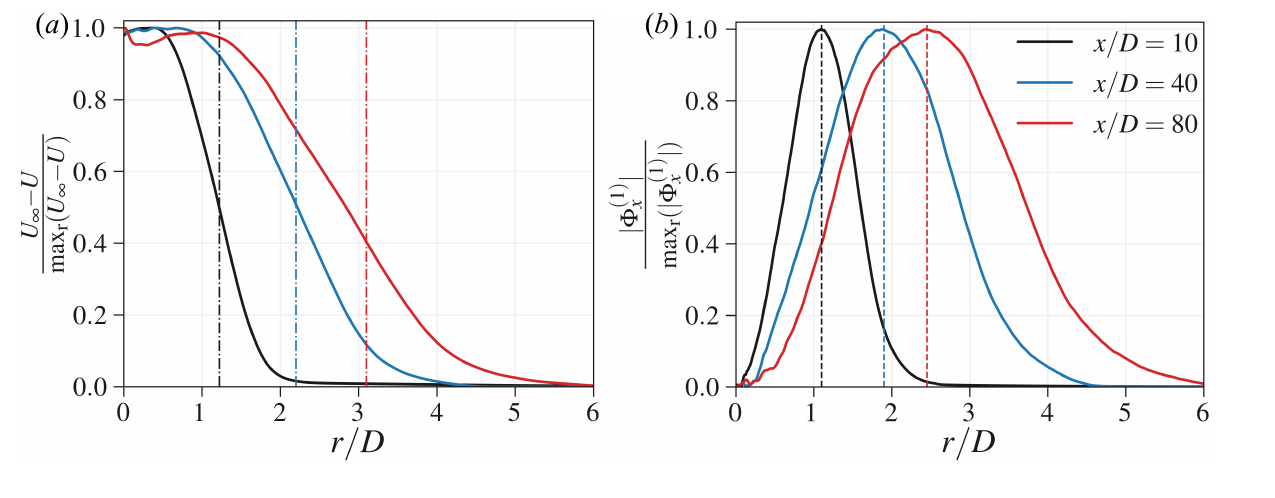}}
\caption{($a$) Normalized mean defect velocity ($U_d = U_\infty - U$) profiles. ($b$) Normalized leading SPOD mode of the streamwise velocity ($u_x$) at $m=2$ and $\St \rightarrow 0$.  Dashed lines in ($a$) and dotted-dashed lines in ($b$) correspond to the radial location of the maximum of SPOD mode and the maximum of mean shear ($\partial U/\partial r$), respectively.}
\label{fig:eigenmode_ud_radial_20_40_80}
\end{figure}

Figure \ref{fig:eigenmode_ud_radial_20_40_80}($a$) shows the normalized radial profiles of mean defect velocity ($U_d$) at $x/D = 10, 40$ and $80$. Figure \ref{fig:eigenmode_ud_radial_20_40_80}($b$) shows the radial profile of the normalized leading SPOD mode's streamwise component for $m=2,\St \rightarrow 0$, at the same streamwise locations as in figure \ref{fig:eigenmode_ud_radial_20_40_80}($a$). As the wake develops in the $x$ direction, the location of mode maximum
 shifts away from the centerline. A visual comparison  shows that the location of amplitude maximum  (dashed lines in figure \ref{fig:eigenmode_ud_radial_20_40_80}$b$) of the dominating streak-containing mode  lies in close proximity to the location of the maximum mean shear (dotted-dashed lines in figure \ref{fig:eigenmode_ud_radial_20_40_80}$a$). The large  radial gradient of the streamwise velocity  induces a positive mean vorticity, lifting up the low-speed fluid from the inner wake  to form streaks. Hence, an extremum in $u'_x$  appears in the SPOD mode. 
 The seminal work of \cite{ellingsen_stability_1975} demonstrates that{, for linearized disturbances in} an inviscid flow, $\partial u_x/\partial t \;  \propto  \; - u_r \partial U(r)/\partial r$, and the lift-up mechanism is most active in the region closest to the largest mean shear. {So is the case in the present turbulent wake.} 


\begin{figure}
\centering
{\includegraphics[trim={0.0cm 2.3cm 0.0cm 2.7cm},clip,width=1.0\textwidth]{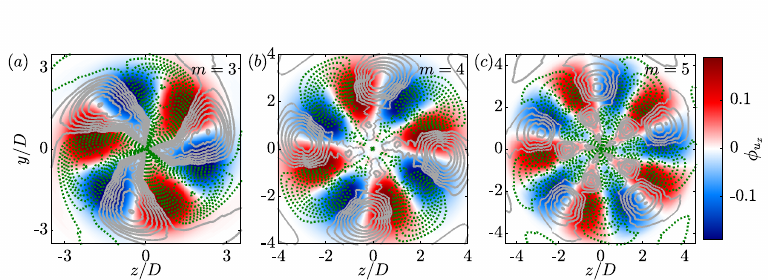}}
\caption{Leading SPOD mode at $x/D=40$ and $St\rightarrow0$ for higher azimuthal wavenumbers: ($a$) $m=3$; ($b$) $m=4$; ($c$) $m=5$. The false colors represent 
$\mathrm{Re}\bqty{\phi^{(1)}_{u_x}(St\rightarrow 0)}$ and the grey and green-dotted lines correspond to the positive and negative  values of streamwise vorticity $\mathrm{Re}\bqty{\phi^{(1)}_{\omega_x}(St\rightarrow 0)}$. }
\label{spod_eigenmode_m_3_4_5}
\end{figure}

{The  lift-up mechanism is also active at higher azimuthal wavenumbers} as demonstrated by figure \ref{spod_eigenmode_m_3_4_5}, which shows the leading SPOD modes at $x/D = 40$ and frequency $\St \rightarrow 0$ for the higher modes: $m = 3$, $4$ and $5$. Similar to figure \ref{fig:spod_eigenmode_20_40_80}, positive and negative streamwise velocity contours are encompassed by counter-rotating vortices that move the fluid from the fast- to slow-speed regions and vice-versa. 
The radial spread of the streamwise velocity lobes increases with $m$ and the number of lobes scales as $2m$. Also, the vortices are shifted by $30^\circ$, $22.5^\circ$, and $18^\circ$, for $m=3$, 4, and 5, respectively. In other words, the set of streamwise velocity lobes for wavenumber $m$ is shifted by an angle of $\pi/2m$ radians with respect to the streamwise vortices. As in the case of $m=2$, the peak of the leading SPOD modes for $m=3,4,5$ lies in the vicinity of the maximum mean shear. Even for higher $m$, lift-up effect occurs nears the region of the largest mean shear. This further confirms that the lift-up mechanism is active for higher azimuthal wavenumbers.

\section{Analysis of triadic interactions in the wake}\label{sec:nonlinear_interaction}

Previous sections show that the streaks are predominantly present in the $m=2$ azimuthal mode. To shed light on the possible dynamics behind the formation of streaks in a turbulent wake, we focus on the nonlinear interactions between the VS-containing $m=1$ mode and the streak-containing $m=2$ mode.

\subsection{Bispectral mode decomposition at select locations}
\label{sec:BMD_results}

\begin{figure}
\centering
{\includegraphics[trim={0.0cm 3.25cm 0.0cm 1.2cm},clip,width=1.0\textwidth]{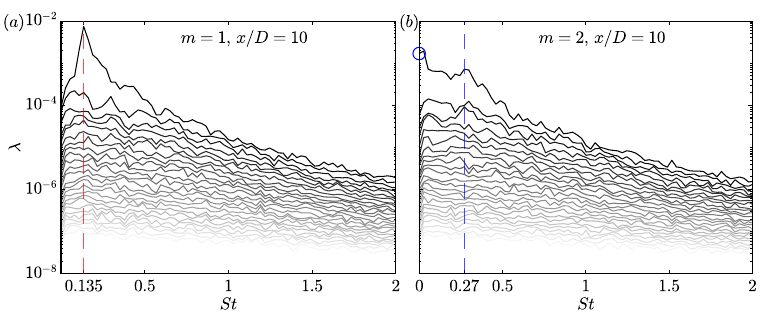}}
\caption{SPOD spectra at $x/D=10$: ($a$) $m=1$; ($b$) $m=2$. Dashed red and blue lines in ($a$) and ($b$) correspond to $\St=0.135$ and $\St=0.27$, respectively.}
\label{fig:SPOD_spectra_x10}
\end{figure}

Figure \ref{fig:SPOD_spectra_x10} shows the SPOD spectra for the azimuthal wavenumbers, $m=1$ and $m=2$, at $x/D =10$. Both SPOD spectra  exhibit a large difference between the first and second eigenvalues for $\St \lesssim 0.5$, thus demonstrating a \emph{low-rank} behaviour. The leading eigenvalue of the $m =1$ azimuthal mode  peaks at the vortex shedding (VS)  frequency, $\St =0.135$. On the other hand, the leading eigenvalue of the $m =2$ azimuthal mode  exhibits a global peak at $\St \rightarrow 0$, and an additional local peak at $\St = 0.27$ (blue dashed line in figure \ref{fig:SPOD_spectra_x10}). 
Furthermore, \citet[see their figure 20]{nidhan2020spectral} find that the VS mode gains prominence at $x/D \approx 1$ while the peak corresponding to $m=2, \St \rightarrow 0$ appears further downstream at $x/D \approx 5$.
These observations collectively point towards different sets of triadic interactions involving the VS mode. 
For example, $m=1, \St=0.135$ can interact with $m=1,\St=-0.135$ to give rise to $m=2, \St\rightarrow 0$ that appears further downstream. Similarly, the self-interaction of  $m=1, \St=0.135$ can generate $m=2, \St=0.27$ (local peak denoted by dashed line in figure \ref{fig:SPOD_spectra_x10}b). 
In what follows, we quantitatively demonstrate that the presence of these triadic interactions at select $x/D$ locations using bispectral mode decomposition (BMD)  \citep{schmidt2020bispectral}.

\begin{figure}
\centering
{\includegraphics[trim={0.0cm 3.7cm 0.0cm 0.2cm},clip,width=1.0\textwidth]{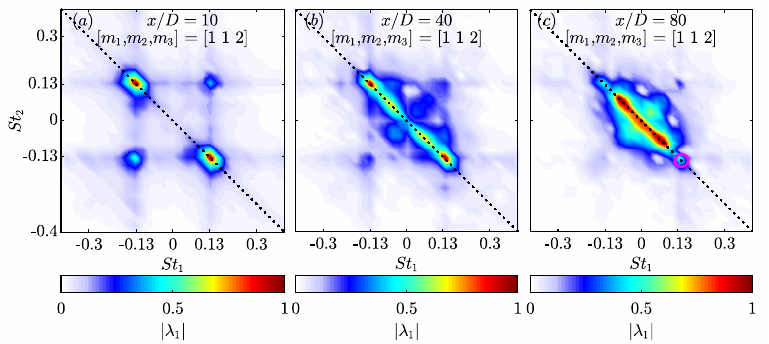}}
\caption{cross-BMD spectra for 
 the azimuthal triad [$m_1$,$m_2$,$m_3 = m_1 + m_2$]=[1,1,2]: ($a$) $x/D=10$, ($b$) $x/D=40$, and ($c$) $x/D=80$. The black dotted line denotes $St_1+St_2 = St_3=0$.}
\label{BMD_spectra_10_40_80}
\end{figure}

Figure \ref{BMD_spectra_10_40_80} shows the cross-mode bispectra for the azimuthal wavenumber triad, [$m_1$,$m_2$,$m_3$] = [1,1,2], at three axial locations, $x/D=10$, 40, and 80. These spectra provide a measure of the interaction among the three azimuthal components. The abscissa ($St_1$) and  ordinate ($St_2$) of the BMD spectra correspond to the frequencies of the $m_1 $ mode and $m_2$ mode, respectively. Here, $m_1 =1$, $m_2 =1$ and $m_3=2$. The high-intensity regions in the spectra represent the energetically  dominant triads. In figure \ref{BMD_spectra_10_40_80}, the mode bispectra are symmetric about the diagonal $St_1=St_2$. At $x/D=10$ (figure \ref{BMD_spectra_10_40_80}($a$)), the most dominant triad is ($0.135, -0.135, 0.0$) and other significant triads are ($0.135, 0.135, 0.27$) and ($-0.135,-0.135,-0.27$). This observation confirms the presence of triadic interactions hypothesized in the context of figure \ref{fig:SPOD_spectra_x10} and emphasizes that the strongest triadic nonlinear interaction is  between the VS mode ($m=1$, $\St=0.135$), its conjugate ($m=1$, $\St=-0.135$), and streaks ($m=2$, $\St \rightarrow 0$). A similar observation is  made from the mode bispectra at $x/D=40$. At $x/D=80$, the most dominant triad occurs at ($0.09,-0.09,0.0$) and along the $\St_1 =-\St_2$ line, but a local maximum (marked by a circle in magenta) is still present at ($0.135,-0.135,0$), which indicates that the triadic interaction between the VS mode and its complex conjugate leading to streaks is still active, albeit subdominant. From figure \ref{BMD_spectra_10_40_80}, we infer that the triadic interaction between [$1,1,2$] at ($0.135,-0.135,0.0$) is prominent in the near wake and its intensity progressively decreases downstream. {It is important to note that when all interactions with  $[m_1,\St_1,m_2,\St_2]$ such that $\St_1 + \St_2 = 0$ were quantified  (not shown here for brevity), that between $m_1=m_2=1$ and $\St_{1,2} = \pm 0.135$ was found to be the strongest}. 
  
\begin{figure}
\centering
{\includegraphics[trim={0.0cm 1.2cm 0.0cm 0.5cm},clip,width=1.0\textwidth]{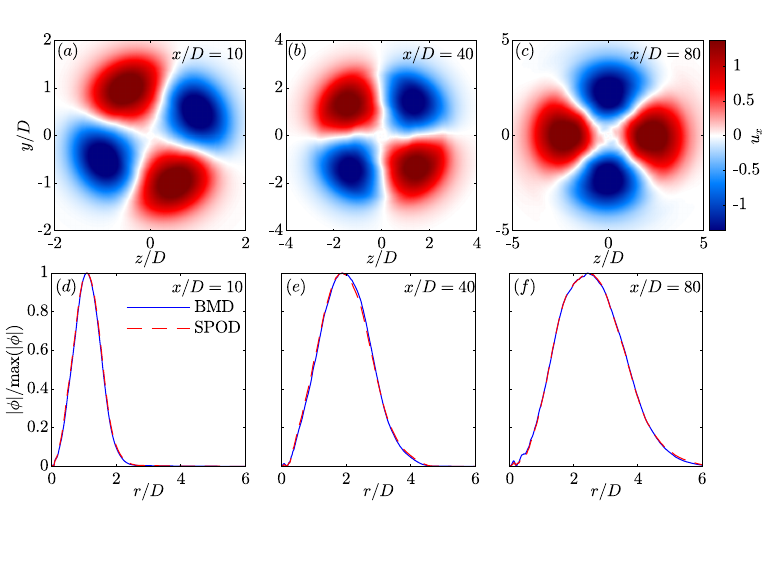}}
\caption{cross-BMD modes for the triad, $[1,1,2]$ in azimuth and $(0.135,-0.135,0.0)$ in frequency,  at three streamwise locations: ($a$) $x/D = 10$; ($b$) $x/D = 40$; ($c$) $x/D = 80$. The false colors represent the real part of streamwise velocity component.The absolute value of the BMD mode and the corresponding leading SPOD mode are compared in ($d$)-($f$). }
\label{BMD_mode_streak_viz_10_40_80}
\end{figure}

Next, we visualize the structures associated with the triadic interaction of the vortex shedding mode, its conjugate, and streaks in figure \ref{BMD_mode_streak_viz_10_40_80}. The real part of the streamwise velocity component of the corresponding cross-bispectral modes is shown at $x/D=10$, 40, and 80. Similar to figure \ref{fig:spod_eigenmode_20_40_80}, the radial extent of these modes increases downstream and they exhibit four lobes of alternate signs. For a more quantitative comparison, figure \ref{BMD_mode_streak_viz_10_40_80} ($d$-$f$) shows the magnitude of the modes normalized by their maximum value. The corresponding curves are coincident, indicating that the spatial structures generated by the triadic interactions are also the most energetic coherent structures. This further confirms that the $m=2,\St \rightarrow 0$ mode is indeed generated through the interaction of $\St = \pm 0.135$ at the $m=1$ azimuthal mode. 

\begin{figure}
\centering
{\includegraphics[trim={0.0cm 1.4cm 0.0cm 1cm},clip,width=1.0\textwidth]{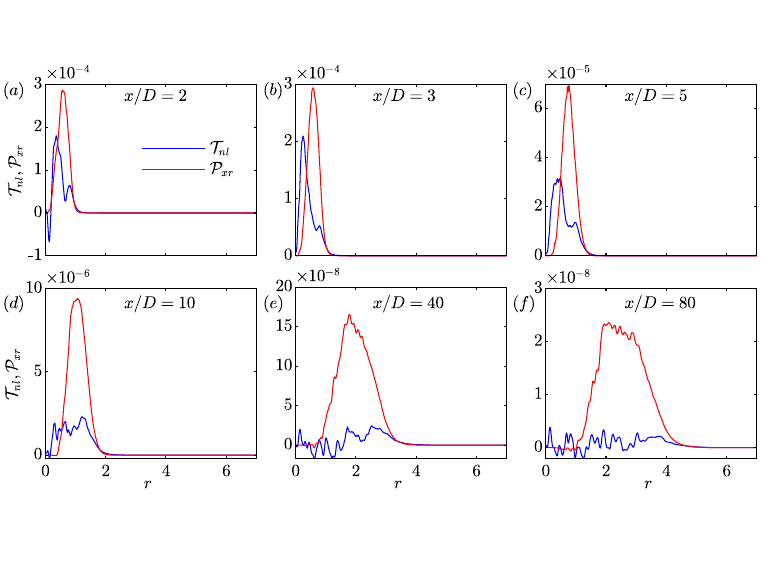}}
\caption{Nonlinear transfer ($\mathcal{T}_{nl}$) from the VS mode and the shear production ($\mathcal{P}_{xr}$) at different streamwise locations for the $m=2,\St \rightarrow 0$ mode: ($a$) $x/D=2$; ($b$) $x/D=3$;  ($c$) $x/D=5$; ($d$) $x/D=10$; ($e$) $x/D=40$; ($f$) $x/D=80$.}
\label{fig_Tnl_P_comp}
\end{figure}

\subsection{Comparison between nonlinear interactions and linear lift-up mechanism}\label{sec:shear_vs_nonlinear}

Finally, the relative role of nonlinear interaction and the linear lift-up mechanism in streak energetics is examined.  In figure \ref{fig_Tnl_P_comp},  $P_{xr}$ denotes shear production in the $m=2,\St \rightarrow 0$ mode and $\mathcal{T}_{nl}$ denotes nonlinear energy transfer from $m=1, \St = \pm 0.135$ modes to the $m=2,\St \rightarrow 0$ mode. These two terms are defined in appendix \ref{sec:appC}. Figure \ref{fig_Tnl_P_comp} reveals that both terms are comparable in magnitude and of the same sign for $x/D \le 5$, whereas $P_{xr}$ dominates beyond $x/D = 10$. Thus, near the wake generator, both (a) the nonlinear interaction of the vortex-shedding mode with its conjugate and (b) the linear production due to the lift-up process are of similar importance to streak energetics. Beyond the near wake, the linear mechanism is responsible for maintaining the streaks.

\section{Discussions and Conclusions} \label{sec:conclusions}

Streaks, which are coherent elongated regions of  streamwise velocity, have been found in a variety of turbulent shear flows. However, they have not received attention in turbulent wakes motivating the present examination of a LES dataset  of flow past a disk at  $\Rey = 50,000$ \citep{chongsiripinyo_decay_2020}. Visualizations and spectral proper orthogonal decomposition (SPOD) are employed and they reveal the presence of streaks from the near wake to the outflow at $x/D \approx 120$. Until now, most of the wake literature has understandably focused on the vortex shedding (VS) structure ($m=1$, $\St = 0.135$ for the circular disk), since  it  is the energetically  dominant coherent structure in  the near and intermediate wake. Upon removing the contribution of the $m=1$ azimuthal wavenumber \textit{a priori} in visualizations, the streaks become evident, even in the near and intermediate wake. Moreover, in the far wake ($x/D \geq 70$), it is the streaks that become the energetically dominant coherent structure. To the best of our knowledge, this is the first study that reports the existence of streaks in turbulent wakes. These results re-emphasize that mean shear, not a wall boundary condition, is a necessary condition for the existence of streaks \citep{jimenez1999autonomous,mizuno2013wall, nogueira_large-scale_2019}.

Streaks differ  from VS structures in three key ways: ($i$) they exhibit a much larger wavelength and time scale; ($ii$) VS structures are tilted with respect to the downstream direction whereas streaks are almost parallel; ($iii$) the  streamwise velocity ($u'_x$)  significantly exceeds  the other two velocity components ($u'_r,u'_\theta$) in magnitude  for streaks whereas $u'_x$ and $u'_r$ are comparable for VS structures. The streaky structures are associated with a frequency of $\St \rightarrow0$ and hence, by Taylor's hypothesis (validated here for the wake) to  a wavenumber of $k_x \rightarrow 0$. While streaky structures are observed for all non-zero azimuthal wavenumbers, $m=2$ dominates in the near-to-far wake. In particular, SPOD analysis reveals that  $m=2$ contains about 55\% (near wake)  to 40\% (far wake)  of the total energy of streaks. This is in contrast to turbulent jets, where \citet{pickering_lift-up_2020}  show that  the dominant azimuthal wavenumber ($m_{dom}$) at $\St \rightarrow 0$ varies as $m_{dom} \sim 1/x$ implying that  higher $m$ and not $m=2 $ would be dominant near the jet nozzle.
It is worth noting that only two studies \citep{johansson_far_2006,nidhan2020spectral} have reported the importance of $m=2$, $\St \rightarrow0$ in turbulent wakes, however they do not link this mode to streaks. 

We find that the lift-up mechanism is active in turbulent wakes, similar to wall-bounded shear flows \citep{abreu_spectral_2020} and turbulent jets \citep{lasagna_near-field_2021,nogueira_large-scale_2019}. Conditional averaging and SPOD analysis clearly demonstrate that  streamwise vortices lift up low-speed fluid from the wake's core and push down high-speed fluid from the outer wake. It is also observed that the lift-up mechanism is spatially most active in the vicinity of the largest mean shear and TKE production,
indicating that energy is directly transferred from the mean flow to the velocity fluctuations in the streaks. 
The lift-up process triggered by the streamwise vortices shows a similar energy transfer mechanism in turbulent pipe flow \citep{hellstrom_self-similarity_2016} and homogeneous shear flow \citep{gualtieri2002scaling,brandt2014lift}.

The lift-up mechanism results in the formation of both low-speed and high-speed streaks. These low- and high-speed streaks exhibit large negative values of Reynolds shear stress. Conditional averaging of streamwise vorticity fluctuations, performed based on peak negative Reynolds shear stress, show that the ejection of low-speed fluid from the wake's core is more dominant than the sweep of the high-speed fluid from the outer wake. The boundary layers also exhibit a similar phenomenon where ejections are a greater contributor to Reynolds shear stress than sweeps \citep{kline1967structure,lu1973measurements}. 

Beyond identification of streaks,  we also explore the role of nonlinear interactions in the context of wake streaks. Specifically, bispectral mode decomposition (BMD) is used to investigate the nonlinear interactions between the $m=1$, $\St= \pm 0.135$  VS mode and the $m=2$, $\St\rightarrow 0$ streak mode. The  $m=1$, $\St= \pm 0.135$ vortices are found to interact and generate the $m=2$, $\St \rightarrow 0$ vortices. These streamwise vortices of the  $m=2$, $\St \rightarrow 0$ mode  then lift up  low-speed fluid from the inner wake and push down the high-speed fluid from the outer wake (figure \ref{fig:spod_eigenmode_20_40_80}) resulting in the formation of streaks. This suggests the wake has a phenomenon analogous to the `regeneration cycle' \citep{hamilton1995regeneration,farrell2012dynamics} of wall-bounded flows, which involves the generation of streamwise vortices through nonlinear interactions and the formation of streaks through linear advection by these streamwise vortices. 
Recently, \citet{bae2021nonlinear} have shown that the nonlinear interactions between spanwise rolls and oblique streaks regenerate streamwise vortices, which then amplify streaks through the lift-up mechanism in wall-bounded flows.  



This work demonstrates that streaks and the associated lift-up mechanism are operative in the turbulent disk wake. The results also open directions for future research. Previous work by \citet{Ortiz2021, ortiz2023high} on the wake of a slender  6:1 prolate spheroid found  that the wake differs significantly from its bluff-body counterpart. Therefore, one possible direction is to investigate how the shape of the wake generator affects the streaky structures and the lift-up mechanism. Moreover, the influence of  the angle of attack and surface properties (roughness, porosity) on the development of these structures could also be explored. Second,  it is worth investigating how density stratification \citep{nidhan2022analysis, gola2023disk}, often found in the natural environment, affects the lift-up mechanism and formation of streaks in turbulent disk wakes.
Lastly, a direct comparison of the characteristics of streaks in wakes, such as length scales, intermittency and life cycle, with those in other turbulent flows such as channel flows and jets can inform us on potentially universal behavior  of streaks in turbulent flows.  In the same vein, it will be also interesting to build reduced-order models to isolate and understand the spatiotemporal features of interaction between the VS mode and streaks as was done with in a problem with KH-like instabilities and streaks \citep{nogueira2021dynamics, cavalieri2022transition}.

\backsection[Acknowledgments]{We thank Dr. K. Chongsiripinyo for the disk wake database.}
\backsection[Funding]{We acknowledge the support of Office  of Naval Research (ONR) grant N00014-20-1-2253.}
\backsection[Declaration of interests] {The authors report no conflict of interest.}
\backsection[Author contributions] {A.N. and S.N. have contributed equally to this paper and are co-first authors.}

\appendix
\section{Effect of parameter, $n_{\rm{fft}}$, on the frequency of the large-scale streaks} \label{sec:appA}
\begin{figure}
\centering
{\includegraphics[trim={0.0cm 1cm 0.0cm 0.9cm},clip,width=1.0\textwidth]{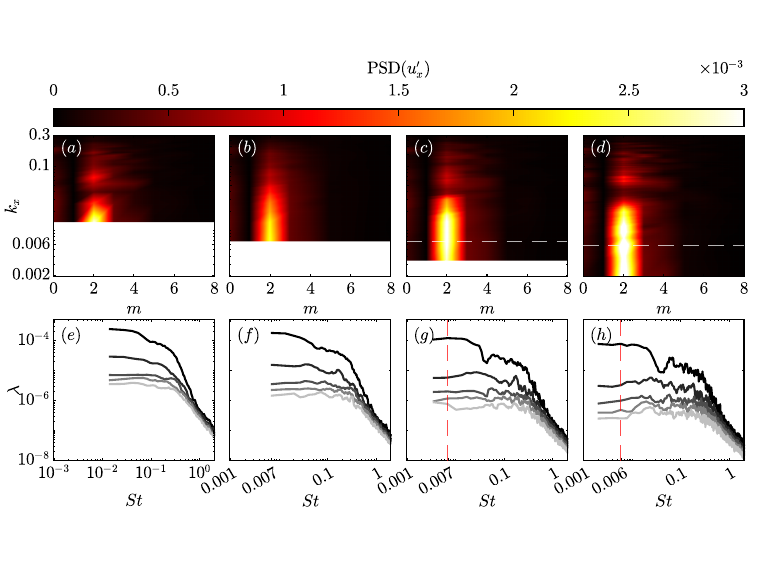}}
\caption{Effect of $n_{\rm{fft}}$ and tapers on the PSD ($a$-$d$) at $r/D=2$, $x/D=40$ and SPOD ($e$-$h$) at $x/D=40$: ($a,e$) $n_{\rm{fft}}=1024$, $N_{\rm{win}}=10$; ($b,f$) $n_{\rm{fft}}=2048$, $N_{\rm{win}}=10$; ($c,g$) $n_{\rm{fft}}=4096$, $N_{\rm{win}}=10$; ($d,h$) $n_{\rm{fft}}=7132$, $N_{\rm{win}}=10$. The white and red lines in the figure correspond to the peak in PSD and SPOD spectra. The PSD and SPOD spectra are plotted for $\St > 0 $. }
\label{N_fft study}
\end{figure}

Here, the parameter $n_{\rm{fft}}$ is varied to identify the \emph{true} frequency of the streaks in the wake.  Due to the constraint of the time-series length, the PSD and SPOD spectra corresponding to the streaks peak in the first frequency bin and are hence interpreted as $\St \rightarrow 0$. For better resolution, one can increase $n_{\rm{fft}}$. However, this results in an increase in variance. Hence, we employ the multitaper-Welch based PSD \citep{thomson1982spectrum} and SPOD \citep{schmidt2022spectral} as it outperforms the standard Welch estimator in terms of variance and resolution \citep{bronez1992performance}.  Recently, \citet{nekkanti2023gappy} have shown that the parameters, $n_{\rm fft}$ and $n_{\rm ovlp}$ can  significantly  affect the spectrum and reconstruction of data. Figure \ref{N_fft study} shows the PSD and SPOD performed using $n_{\rm{fft}} = 1024$, $2048$, $4096$ and $7132$ with $N_{\rm{win}} = 10$ Slepian tapers as windows. The PSD is computed at $x/D = 40, r/D=2$ and SPOD spectra at $x/D = 40$. As the goal is to identify the true frequency, we plot the spectra for $\St > 0 $. For figure \ref{N_fft study} ($a$,$b$,$e$,$f$), as the $n_{\rm fft}$ is not large enough, the peak in the PSD and SPOD is at $\St \rightarrow 0$. On increasing the $n_{\rm{fft}}$, two-dimensional PSD and SPOD spectra can resolve the lower frequencies and their peak approaches $\St \approx 0.006$ at the highest possible $n_{\rm{fft}}$ (figure \ref{N_fft study}$d,h$). Using this approach, the frequency associated with the large-scale streaky structures can be identified, which is $St\approx 0.006$. Note that, the availability of more snapshots will result in a better convergence of the peak frequency associated with streaks.

\section{Conditional averaging: sensitivity of parameter, $c$} \label{sec:appB}

\begin{figure}
\centering
{\includegraphics[trim={0.0cm 9cm 0.0cm, 0cm},clip=true,width=0.8\linewidth]{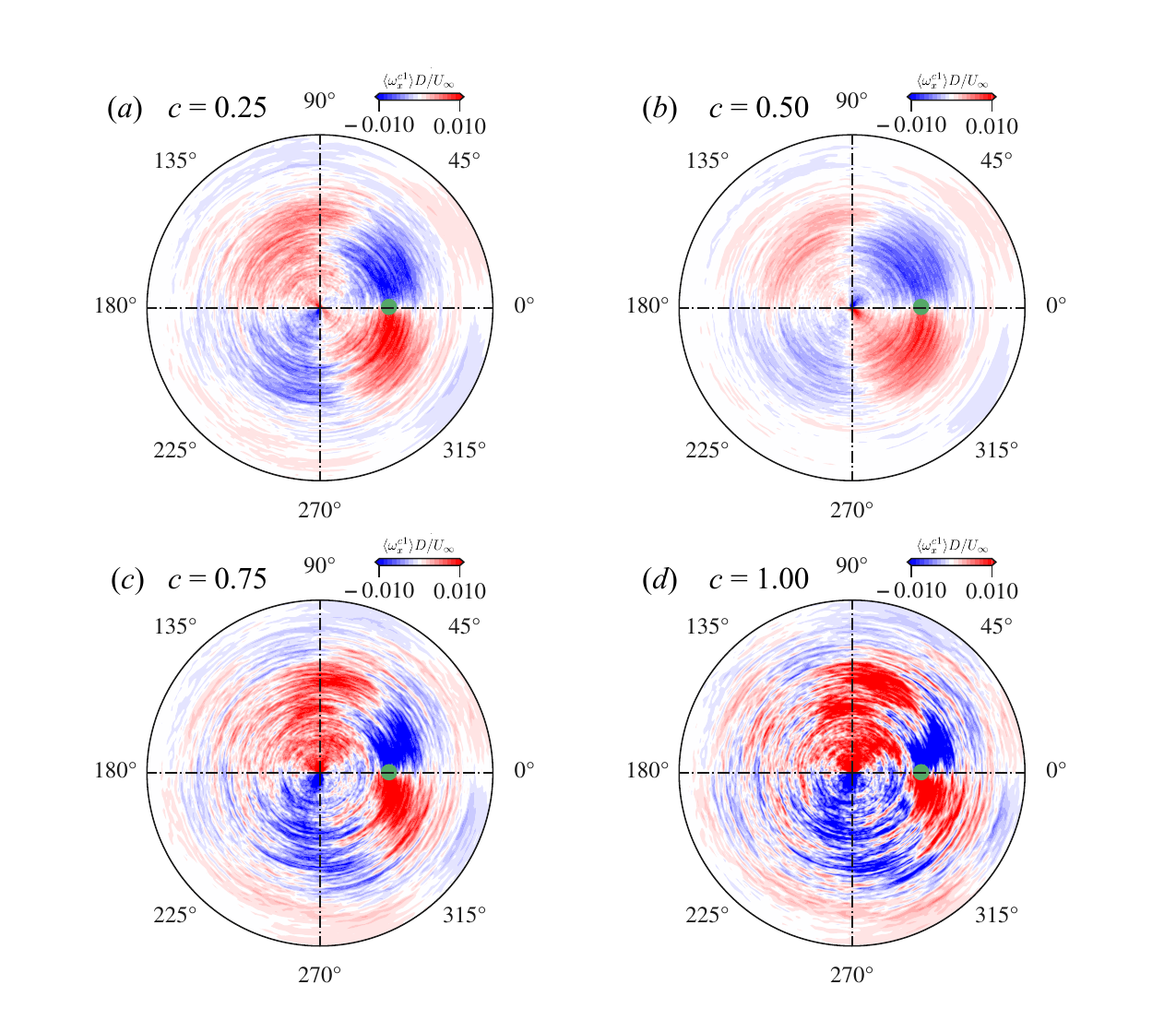}}
\caption{Effect of changing $c$ on the conditionally averaged streamwise vorticity $\langle \omega_x^{c1} \rangle$ obtained from $u'_x$ based conditioning at $x/D = 40$.}
\label{fig:cond_average_appendix_uxur}
\end{figure}

\begin{figure}
\centering
{\includegraphics[trim={0.0cm 9cm 0.0cm, 0cm},clip=true,width=0.8\linewidth]{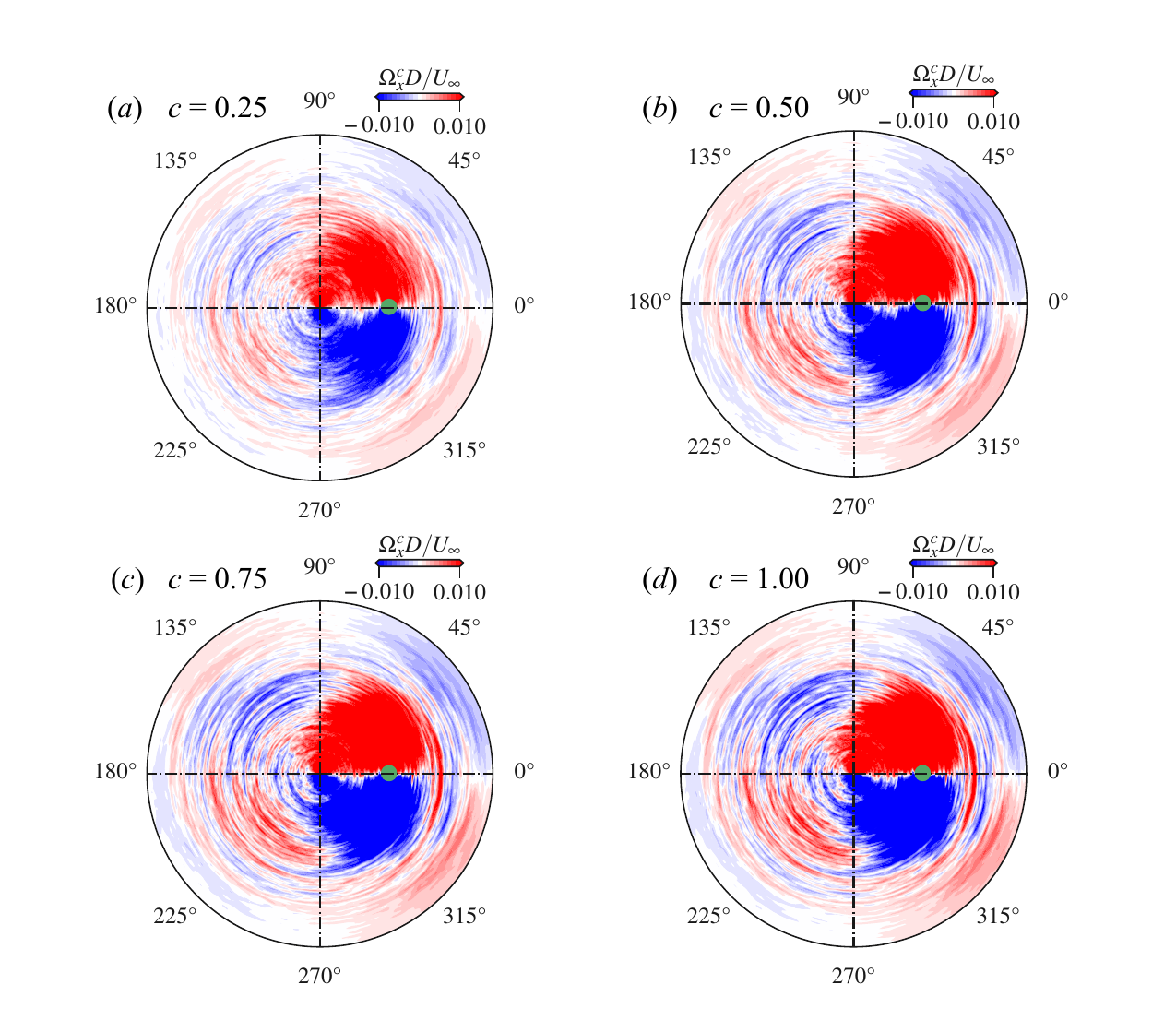}}
\caption{Effect of changing $c$ on the conditionally averaged streamwise vorticity $\langle \omega_x^{c2} \rangle$ obtained from $u'_x u'_r$ based conditioning at $x/D = 40$.}
\label{fig:cond_average_appendix_ux}
\end{figure}

Figures \ref{fig:cond_average_appendix_uxur} and \ref{fig:cond_average_appendix_ux} show the effect of varying $c$ in equations \ref{eq:ux_condition} and \ref{eq:uxur_condition} on the obtained $\langle \omega_x^{c1}\rangle$ and $\langle \omega_x^{c2}\rangle$ fields, respectively, at a representative location of $x/D = 40$. In both figures, changing $c$ between 0 to 1 has little qualitative effect on the distribution of negative and positive vortices around the conditioning point. It is interesting to note, however, that  the  $\langle\omega_{x}^{c2}\rangle$ field is more robust to changes in $c$ than the $\langle\omega_{x}^{c1}\rangle$ field. Increasing the value of $c$ leads to the capture of more intense events; but, it also reduces the number of realizations available for averaging. Hence, based on figure \ref{fig:conditional_avg_omegax}, $c=0.5$ is selected as a compromise between capturing intense events and having sufficient realizations for temporal averaging. 

\section{Scale specific production and non-linear transfer terms in the disk wake for $m=2, \St \rightarrow 0$ mode} \label{sec:appC}

Fourier modes and Welch estimation are employed to compute the scale-specific production and nonlinear transfer terms. The mathematical form of these two terms is as follows.

The scale-specific production term is 
\begin{equation}
     \mathcal{P}\pqty{m, \St} =  -\mathrm{Re}\bqty{\hat{u}_j^*\pqty{m,\St}\hat{u}_i \pqty{m,\St} \frac{\partial U_j}{\partial x_i}},
\end{equation}
where, $\mathrm{Re}\bqty{\cdot}$ denotes the real part. We are specifically interested in the production term by lateral mean shear operating on  the $m=2,\St\rightarrow0$ mode:

 \begin{equation}
     \mathcal{P}_{xr}\pqty{m=2, \St \rightarrow 0} =  -\mathrm{Re}\bqty{\hat{u}_x^*\hat{u}_r\frac{\partial U}{\partial r}}_{\pqty{m=2, \St \rightarrow 0}}  .
 \end{equation}

The scale-specific nonlinear transfer is 
 \begin{equation}
     \mathcal{T}_{nl}\pqty{m, \St} =  - \mathrm{Re}\bqty{\hat{u}_j^*\pqty{m, \St}\widehat{u_i\frac{\partial u_j}{\partial x_i}}\pqty{m, \St}}
 \end{equation}
 
This term represents the total nonlinear transfer term that generates the frequency $\St$ and wavenumber $m$. For identifying the triadic energy transfer, i.e., $m_1 +m_2 =m_3$ and $\St_1 + \St_2 =\St_3$, following \cite{cho2018scale}, we write this term in  discretized convolution:

\begin{equation}
     \mathcal{T}_{nl}\pqty{m_3, \St_3} =  - \mathrm{Re}\bqty{{\hat{u}_j^*\pqty{m_3, \St_3}\sum\limits_{\substack{\St_1 +\St_2=\St_3 \\ m_1 +m_2=m_3}}\hat{u}_i\pqty{m_1, \St_1}\frac{\partial \hat{u}_j}{\partial x_i}\pqty{m_2, \St_2}}}
\end{equation}.

For a single triad [$m_1,m_2,m_3$], [$\St_1,\St_2,\St_3$] this is expanded as follows:

\begin{equation}
     \mathcal{T}_{nl}\bqty{(m_1,m_2,m_3), (\St_1,\St_2,\St_3)} =  - \mathrm{Re}\bqty{{\hat{u}_j^*\pqty{m_3, \St_3}\hat{u}_i\pqty{m_1, \St_1}\frac{\partial \hat{u}_j}{\partial x_i}\pqty{m_2, \St_2}}}
\end{equation}

The nonlinear term of our interest is the one responsible for transfer of energy from the VS mode to the $m=2,\St\rightarrow 0$ mode:

\begin{equation}
     \mathcal{T}_{nl}\bqty{(1,1,2), (0.135,-0.135,0)} =  - \mathrm{Re}\bqty{{\hat{u}_j^*\pqty{2,0}\hat{u}_i\pqty{1, 0.135}\frac{\partial \hat{u}_j}{\partial x_i}\pqty{1, -0.135}}}.
\end{equation}



\bibliographystyle{jfm}
\bibliography{streaks_re5e4_frinf}
\end{document}